\documentclass[magazine]{IEEEtran}
\usepackage{graphicx}
\usepackage{color}
\usepackage[noend]{algpseudocode}
\usepackage{placeins}
\usepackage{float}
\usepackage{tabularx,colortbl}
\usepackage{amsthm}
\usepackage{amsmath}
\usepackage{mathtools}
\usepackage{amssymb}
\usepackage{cleveref}
\usepackage{algorithm}
\usepackage{verbatim}
\usepackage{url}
\usepackage{cases}
\usepackage{amsmath}
\usepackage{amsfonts}
\usepackage{amssymb}
\usepackage{verbatim}
\usepackage{multirow}
\usepackage{graphicx}
\usepackage{mathtools}
\usepackage{multicol}
\usepackage{epstopdf}
\usepackage{subfigure}
\usepackage{graphics}
\usepackage{color}
\usepackage{xcolor}
\usepackage{MnSymbol}
\usepackage[all,cmtip]{xy}
\definecolor{BoxBackground}{RGB}{235,235,255}

\IEEEoverridecommandlockouts

\begin{document}
\title{Sub-Nyquist Radar Systems:\\ Temporal, Spectral and Spatial Compression}
\author{

%\thanks{
%This work was funded by the European Union's Horizon 2020 research and innovation program under grant agreement ERC-BNYQ, and by the Israel Science Foundation under Grant no. 335/14. Deborah Cohen is grateful to the Azrieli Foundation for the award of an Azrieli Fellowship.}
Deborah Cohen and Yonina C. Eldar \\
Technion - Israel Institute of Technology, Haifa, Israel \\
\{debby@campus, yonina@ee\}.technion.ac.il}

\maketitle

% Some of my macros (SHAHAR)
\global\long\def\dint#1#2#3#4{\int\limits _{#1}^{#2}#3\,\text{d}#4}
\global\long\def\v#1{\mathbf{#1}}
\global\long\def\sinc{\text{sinc}}
\global\long\def\hz#1{\,\text{#1Hz}}

\section{Introduction}

Radar is an acronym for "radio detection and ranging". However, the functions of today's radar systems, both in civilian and military applications, go beyond simple target detection and localization. They extend to tracking, imaging, classification and more, and involve different types of radar systems such as through-the-wall radar \cite{masbernat2010mimo}, ground penetration radar \cite{daniels2004ground}, automotive radar \cite{schuler2005array} and weather radar \cite{bringi2001polarimetric}. Although radar technology has been well established for decades, a new line of compressed radars has recently emerged. These aim at reducing the complexity of classic radar systems by exploiting inherent prior information on the structure of the received signal from the targets. The goal of this paper is to review these novel sub-Nyquist radars and their potential applications.

Conventional radar systems transmit electromagnetic waves of near-constant power in very short pulses towards the targets of interest. Between outgoing pulses, the radar measures the signal reflected from the targets to determine their presence, range, velocity and other characteristics. Different systems use different radar waveforms and varying transmit strategies. One of the most popular methods is pulse-Doppler radar which periodically transmits identical pulses. In contrast, stepped frequency radars (SFR)~\cite{skolnik} vary the carrier frequency of each pulse. Some systems rely on simple traditional waveforms such as Gaussian pulses while others adopt more complex signals, such as chirps~\cite{peebles2007radar, richards2005fundamentals}. Each configuration corresponds to a certain choice in the complexity-performance trade-off, between complex waveform and system designs and target detection and estimation.

State of the art radar systems operate with large bandwidths, large coherent processing intervals (CPIs) and high number of antennas in multiple input multiple output (MIMO) settings~\cite{fishler2004mimo, li2009mimo}, in order to achieve high range, velocity and azimuth resolution, respectively. This, in turn, generates large data sets to be sampled, stored and processed, creating a bottleneck both in terms of analog system complexity, including high rate analog-to-digital converters (ADCs), and subsequent digital processing~\cite{SamplingBook}.

In the past few years, novel approaches to radar signal processing have emerged that allow radar signal detection and parameter estimation using a much smaller number of measurements than that required by spatial and temporal Nyquist sampling. While temporal sampling refers to taking samples in time intervals determined by the sampling rate, spatial sampling extends this notion to placing transmit and receive antennas whose locations are governed by the signal wavelength. These works capitalize on the fact that, in most radar applications, the reflectivity scene consists of a small number of strong targets. That is, the reflected signals by only a few targets have high enough power to be detected by the radar receiver. In pulse-Doppler radar, the target scene is often sparse in the joint time-frequency, or ambiguity, domain \cite{skolnik}. In synthetic aperture radar (SAR) \cite{curlander1991synthetic}, the scene is often sparse in the Fourier or wavelet domain, or even in the image domain. 

Over the past decade, many works have exploited the inherent sparsity of the target scene to enhance radar estimation capabilities. These rely on the compressed sensing (CS) \cite{CSBook, SamplingBook} framework, brought to the forefront  by the works of Candes, Romberg and Tao \cite{candes2006robust, candes2006stable}, and of Donoho \cite{donoho2006compressed}. Although the natural application of CS is typically the reduction of the required number of samples to perform a certain signal processing task, it was first used by the radar community to increase a target's parameter resolution~\cite{herman2009high, chi2009golay, shah2009step, strohmersparse, strohmersparse2, gedalyahu2011identification}. It was later applied to reduce the number of samples to be processed~\cite{teke2014robust, ender2010compressive, yu2010mimo, kalogerias2014matrix, mimoMC} and finally to reduce the sampling rate~\cite{ertin2011sparse, liu2015pulse} and number of antennas~\cite{rossi2014spatial} required in radar systems, performing time and spatial compression and alleviating the burden on both the analog and digital sides. In particular, the recently proposed Xampling (``compressed sampling") concept \cite{SamplingBook, mishali2011xampling}, has been applied to radar \cite{barilan2014focusing, cohen2016towards, cohen2016summer} in order to break the link between bandwidth, CPI and number of antennas on the one hand and range, Doppler and azimuth resolution, respectively, on the other.

Reviews of compressive radar~\cite{ender2010compressive, potter2010sparsity, cetin2014sparsity, zhao2016race} mostly deal with radar imaging. The works~\cite{potter2010sparsity, cetin2014sparsity} focus on SAR imaging and consider sparsity based radar imagery using both greedy algorithms, that iteratively recover the sparse target scene, and convex relaxations of sparsity inducing regularization. The special cases of interferometric, polarimetric and circular SAR are presented in \cite{potter2010sparsity} for both 2D and 3D images. In~\cite{cetin2014sparsity}, diverse SAR applications are reviewed, such as wide-angle SAR imaging, joint imaging and autofocusing from data with phase errors, moving targets, analysis and design of SAR sensing missions. A survey of statistical sparsity based techniques for radar imagery applications is presented in~\cite{zhao2016race}, including super-resolution imaging, enhanced target imaging, auto-focusing and moving target imaging. The review of~\cite{ender2010compressive} presents three applications of CS radars: pulse compression, radar imaging and air space surveillance with array antennas. At the time it was written, there were only a small number of publications addressing the application of CS to radar, as stated by the authors.

In this article, we focus on non radar imaging applications and survey many recent works that exploit CS in different radar systems, to achieve various goals. We consider different transmit waveforms and processing approaches, while focusing on pulse-Doppler radar, which is one of the most popular systems, and its extension to MIMO configurations. Our goal is to review the main impacts of compressed radar on parameter resolution as well as digital and analog complexity. The survey includes fast time compression schemes, which reduce the number of acquired samples per pulse, slow time compression techniques, that decrease the number of pulses, and spatial compression approaches, where the number of transmit and receive antenna elements is reduced. We show that beyond substantial rate reduction, compression may also enable communication and radar spectrum sharing~\cite{griffiths2015radar, fitz2014towards, bernhard2010final}, as elaborated on in~\cite{cohen2016specx}. Throughout the paper, we consider both theoretical and practical aspects of compressed radar, and present hardware prototype implementations \cite{baransky2014prototype, mishra2016cognitive, cohen2016radarconmimo, yoo2012compressed} of the theoretical concepts, demonstrating real-time targets' parameters recovery from low rate samples in pulse-Doppler and MIMO radars.

\section{Radar Systems}

Radar systems aim at estimating targets' parameters to determine their location and motion. In its simplest form, the radar transmits a single pulse towards targets in one direction and recovers their range, i.e., distance to the radar, which is proportional to the received pulse delay. More elaborate systems are able to provide additional information on the targets. Pulse-Doppler radars transmit several pulses, enabling them to resolve both the targets' ranges and radial velocities, which is proportional to the Doppler frequency. Stepped frequency based approaches achieve high effective bandwidth to increase range resolution, while allowing for narrow instantaneous bandwidth. MIMO radars use several elements both at the transmitter and at the receiver to illuminate the entire target scene and recover targets' azimuths in addition to their range and velocity. In this review, we consider the application of compression in terms of the number of required samples, pulses and antennas, and its impact on different aspects of the radar system, including parameter resolution and system complexity, for several types of radars.

\subsection{Pulse-Doppler Radar}

%%%%%%%%%%%%%%%%%%%%%%%%%%%%%%%%%%%%%%%%
% The Targets BOX
%%%%%%%%%%%%%%%%%%%%%%%%%%%%%%%%%%%%%%%%
\begin{figure*}[tb]\fboxsep1em
\colorbox{BoxBackground}{\begin{minipage}{1\textwidth}\begin{multicols*}{2}
\section*{Targets' Assumptions}
\label{box:target}

In order to simplify the received signal model, the following assumptions on the targets' locations and motions are typically made \cite{skolnik}:
\begin{itemize}
\item[\textbf{A1}] Far targets - target-radar distance is large compared with the distance change during the CPI, which allows for constant $\alpha_l$ within the CPI:
\begin{equation}
\dot{r}_l P \tau \ll r_l \Rightarrow \nu_l \ll \frac{f_c \tau_l}{P \tau}.
\end{equation}
\item[\textbf{A2}] Slow targets - constant Doppler phase during pulse time:
\begin{equation}
\nu_l T_p \ll 1,
\end{equation}
and low target velocity allows for constant $\tau_l$ during the CPI. This condition holds when the baseband Doppler frequency is smaller than the frequency resolution:
\begin{equation}
\frac{2 \dot{r}_l B_h}{c} \ll \frac{1}{P\tau} \Rightarrow \nu_l \ll \frac{f_c}{P \tau B_h}.
\end{equation}
\item[\textbf{A3}] Small acceleration - target velocity remains approximately constant during the CPI allowing for constant $\nu_l$. This condition is satisfied when the velocity change induced by acceleration is smaller than the velocity resolution:
\begin{equation}
\ddot{r}_l P \tau \ll \frac{c}{2 f_c P \tau} \Rightarrow \ddot{r}_l \ll \frac{c}{2 f_c (P\tau)^2}.
\end{equation}
\end{itemize}
Although these assumptions may seem hard to comply with, they all rely on slow enough relative motion between the radar and its targets. Radar systems tracking people, ground vehicles, and sea vessels usually comply quite easily \cite{peebles2007radar}.

In MIMO settings, two additional assumptions are adopted on the array structure and transmitted waveforms:
\begin{itemize}
\item[\textbf{A4}] Collocated array - target RCS $\alpha_l$ and $\theta_l$ are constant over the array (see \cite{haim2006} for more details).
\item[\textbf{A5}] Narrowband waveform - small aperture allows $\tau_l$ to be constant over the channels,  
\begin{equation} \label{eq:A1}
\frac{2Z \lambda}{c} \ll \frac{1}{B_h}.
\end{equation}
\end{itemize}
\end{multicols*}
\end{minipage}}\end{figure*}

\begin{figure}
\begin{center}
\includegraphics[width=1\columnwidth]{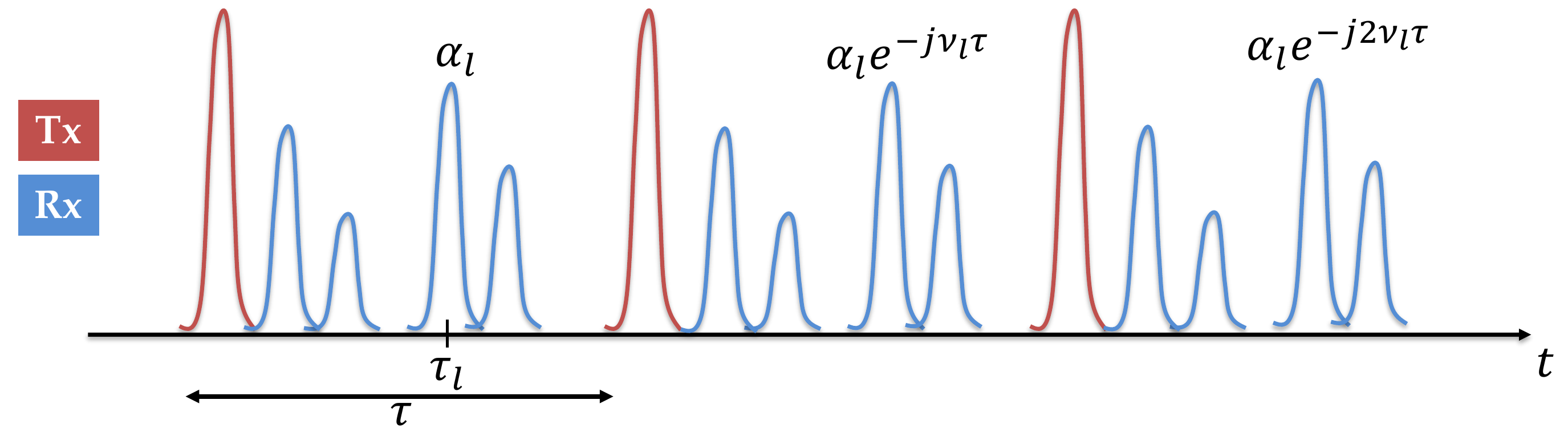}
\caption{Pulse-Doppler radar transmitted (Tx) and received (Rx) pulse train with $P=3$ pulses and $L=4$ targets.}
\label{fig:pulses}
\end{center}
\end{figure}

A standard pulse-Doppler radar transceiver detects targets by transmitting a periodic stream of pulses and processing its reflections. The transmitted signal $x_T(t)$ consists of $P$ equally spaced pulses $h(t)$ such that
\begin{equation}
\label{eq:uni_model}
x_T(t)= \sum_{p=0}^{P-1} h(t-p\tau), \quad 0 \leq t \leq P \tau.
\end{equation}
The pulse-to-pulse delay $\tau$ is the pulse repetition interval (PRI), and its reciprocal $1/\tau$ is the pulse repetition frequency (PRF). The entire span of the signal in (\ref{eq:uni_model}), namely $P\tau$, is the CPI. The pulse time support is denoted by $T_p$, with $0 < T_p < \tau$. The pulse $h(t)$ is typically a known time-limited baseband function with continuous-time Fourier transform (CTFT) $H(f)=\int_{- \infty}^{\infty} h(t) e^{-j 2\pi f t} dt$ that has negligible energy at frequencies beyond $B_h/2$, where $B_h$ is referred to as the bandwidth of $h(t)$. An example of a transmitted pulse train is illustrated in Fig.~\ref{fig:pulses}.

It is typically assumed that the target scene is composed of $L$ non-fluctuating point-targets, according to the Swerling-0 model \cite{skolnik}. This is one of the popular models in the radar signal processing literature since, by describing an idealized target, it allows simplifying the radar equations while constituting a fairly good approximation in many applications \cite{peebles2007radar, richards2005fundamentals}. Other models, such as Swerling-1, which applies to targets composed of many independent scatters, or fluctuating target models, are beyond the scope of this paper. The pulses reflect off the $L$ targets and propagate back to the transceiver. Each target $l$ is defined by three parameters: 
\begin{itemize}
\item A time delay $\tau_l=2r_l/c$, proportional to the target's distance to the radar or range $r_l$, where $c$ is the speed of light.
\item A Doppler radial frequency $\nu_l=2\dot{r}_lf_c/c$, proportional to the target's radial velocity to the radar, namely the target's velocity radial component $\dot{r}_l$, and the radar's carrier frequency $f_c$.
\item A complex amplitude $\alpha_l$, proportional to the target's radar cross section (RCS), dispersion attenuation and other propagation factors. 
\end{itemize}
The targets are defined in the radar radial coordinate system and are typically assumed to lie in the radar unambiguous time-frequency region: delays up to the PRI and Doppler frequencies up to the PRF. When this assumption does not hold, several processing techniques have been proposed that require the transmission of multiple pulse trains with different parameters, e.g., different PRFs. We review this setting in the ``Range-Velocity Ambiguity Resolution" section.

Based on the three assumptions \textbf{A1-A3} presented in ``Targets' Assumptions", the received signal can be written as
\begin{equation}
\label{eq:uni_rec}
x_R(t)= \sum_{p=0}^{P-1} \sum_{l=0}^{L-1} \alpha_l h(t-\tau_l - p\tau) e^{-j \nu_l p \tau}, \quad 0 \leq t \leq P\tau.
\end{equation}
It will be convenient to express $x_R(t)$ as a sum of single frames
\begin{equation}
\label{eq:frames}
x_R(t)= \sum_{p=0}^{P-1} x_p(t),
\end{equation}
where
\begin{equation}
\label{eq:one_frame}
x_p(t)= \sum_{l=0}^{L-1} \alpha_l h(t-\tau_l - p\tau) e^{-j \nu_l p \tau}, \quad 0 \leq t \leq P \tau.
\end{equation}
An illustration of a received pulse train is shown in Fig.~\ref{fig:pulses} with $L=4$ targets.
In pulse-Doppler radar, the goal is to recover the $3L$ parameters $\{\tau_l, \nu_l, \alpha_l\}$, for $0 \leq l \leq L-1$ from the received signal $x_R(t)$. In particular, estimating the time delays $\tau_l$ and Doppler frequencies $\nu_l$ enables approximation of the targets' distances and radial velocities.

\subsection{Stepped Radar Waveforms}

In classic pulse-Doppler radar, high range resolution requires large signal bandwidth. This technology bottleneck is partially overcome by step frequency based waveforms, in which the large bandwidth is obtained sequentially by stepping the frequency of each pulse, keeping the instantaneous bandwidth low. Two popular examples of such waveforms are SFR and stepped chirps. An SFR~\cite{skolnik} system transmits $P$ narrowband pulses, where each pulse $p$ has carrier frequency
\begin{equation}
f_p=f_0+p\Delta_f,
\end{equation}
for $0 \leq p \leq P-1$, with $f_0$ the initial frequency and  $\Delta f$ the frequency increment. The $p$th transmitted pulse is a rectangular pulse modulated by its carrier $f_p$. The corresponding received signal is then of the form
\begin{equation}
\label{eq:sfr}
x_p(t)=\sum_{l=0}^{L-1} \alpha_l \text{rect}(t-\tau_l) e^{-j 2 \pi f_p(t-\tau_l)}e^{j\nu_lp\tau}.
\end{equation}

To process the received signal, the delay is neglected in the signal envelope due to the narrowband assumption. An SFR traditionally obtains one sample from each received pulse and computes the phase detector output sequence as
\begin{equation}
\label{eq:sfr_phase_detector}
y_p=\sum_{l=0}^{L-1} \alpha_l e^{j 2 \pi f_p\tau_l}e^{j\nu_lp\tau}.
\end{equation}
The phase detector signal $y_p$ can be modeled as the product of the received signal (\ref{eq:sfr}) and the reference signal, followed by a low-pass filter (LPF). Conventional processing applies an inverse discrete Fourier transform (DFT) on the output in order to estimate the targets' time delays $\tau_l$ and Doppler frequencies $\nu_l$. The range resolution achieved by SFR is $\frac{c}{2P \Delta_f}$, where $P\Delta_f$ is the total effective bandwidth of the signal over $P$ pulses.

Another popular stepped waveform is the stepped chirp or multi-frequency chirp signal. The corresponding transmitted signal is given by
\begin{equation} \label{eq:trans_chirp}
x_T(t)=\sum_{p=0}^{P-1} e^{j \phi_p} \text{rect}\left( \frac{t}{\tau} \right) e^{j 2 \pi (f_pt+\frac{\gamma}{2}t^2)},
\end{equation}
where $\gamma$ is the common chirp rate and $f_p$ and $\phi_p$ are the frequency and complex phase of the $p$th sub-carrier. The returned signal corresponding to the $p$th pulse, given by
\begin{equation}
x_p(t)=\sum_{l=0}^{L-1} \alpha_l e^{j \phi_p} \text{rect}\left( \frac{t-\tau_l}{\tau} \right) e^{j 2 \pi (f_p (t-\tau_l)+\frac{\gamma}{2}(t-\tau_l)^2)},
\end{equation}
is dechirped with a reference linear frequency waveform of fixed frequency equal to the first carrier $f_0$:
\begin{equation}
m(t) = \text{rect}\left( \frac{t-\tau_r}{\tau} \right) e^{-j 2 \pi (f_0 t +\frac{\gamma}{2}t^2)}.
\end{equation}
The receive window is $\tau_r = 2(r_{\text{max}}+r_{\text{min}})/c$ and the reference delay is $t_r=(r_{\text{max}}+r_{\text{min}})/c$, with $r_{\text{max}}$ and $r_{\text{min}}$ the maximal and minimal ranges, respectively. The resulting dechirped received signal can be written as
\begin{equation} \label{eq:chirp_sig}
x_p(t)=\sum_{l=0}^{L-1} \alpha_l e^{j (\phi_p-2\pi f_p \tau_l)} \text{rect}\left( \frac{t -\tau_l+t_r}{\tau} \right) e^{j 2 \pi (f_p-f_0 -\gamma \tau_l) t}.
\end{equation}
Classic processing of the received signal includes a DFT operation to recover the targets' delays $\tau_l$.

%%%%%%%%%%%%%%%%%%%%%%%%%%%%%%%%%%%%%%%%
% The STANDARD PROCESSING BOX
%%%%%%%%%%%%%%%%%%%%%%%%%%%%%%%%%%%%%%%%
\begin{figure*}[h!]\fboxsep1em
\colorbox{BoxBackground}{\begin{minipage}{1\textwidth}\begin{multicols*}{2}
\section*{Classic pulse-Doppler and MIMO Processing}
\label{box:classic_process}

Classic methods for radar processing typically consist of the following stages \cite{skolnik, CookBook}:
\begin{enumerate}
\item {\bf Sampling}: Sample each incoming frame $x_p(t)$ at its Nyquist rate $B_h$, equal to the double-sided bandwidth of $h(t)$, creating the samples $x_p[n], 0 \leq n \leq N-1$, where $N=\tau B_h$. We assume for simplicity that $N$ is an integer.
\item {\bf Matched filter}: Apply a standard matched filter (MF) on each frame $x_p[n]$. This results in the outputs $y_p[n]=x_p[n] \ast h[-n]$, where $h[n]$ is the sampled version of the transmitted pulse $h(t)$, at its Nyquist rate and $\ast$ is the convolution operation. The time resolution attained in this step is $1/B_h$. 
\item {\bf Doppler processing}: For each discrete time $n$, perform a $P$-point DFT along the pulse dimension, namely $z_n[k]=\text{DFT}_P\{y_p[n]\}=\sum_{p=0}^{P-1} y_p[n]e^{-j2\pi pk/P}$ for $0 \leq k \leq P$. The Doppler resolution is $1/P\tau$.
\item {\bf Delay-Doppler map}: Stacking the vectors $\mathbf{z}_n$, and taking absolute value, we obtain a delay-Doppler map $\mathbf{Z}=\text{abs}[\mathbf{z}_0, \dots, \mathbf{z}_{N-1}] \in \mathbb{R}^{P \times N}$.
\item {\bf Peak detection}: A heuristic detection process, where knowledge of number of targets, targets' powers, clutter location, etc. may help in discovering targets' positions. For example, if we know there are $L$ targets, then we can choose the $L$ strongest points in the map. Alternatively, constant false alarm rate (CFAR) detectors determine the power threshold above which a peak is considered to originate from a target, so that a required probability of false alarm (FA) is achieved \cite{scharf1991statistical, gandhi1988analysis}.
\end{enumerate}

Classic collocated MIMO radar processing traditionally includes the following stages:
\begin{enumerate}
  \item \textbf{Sampling:} At each receiver $0 \leq q \leq R-1$, where $R$ denotes the number of receivers, the signal $x_q(t)$ is sampled at its Nyquist rate $B_{\text{tot}}$. In code division multiple access (CDMA) and time division multiple access (TDMA), $B_{\text{tot}}=B_h$ as all waveforms overlap in frequency, whereas in frequency division multiple access (FDMA), $B_{\text{tot}}=TB_h$, where $B_h$ denotes the bandwidth of a single waveform in both cases and $T$ is the number of transmitters.
  \item \textbf{Matched filter:} The sampled signal is convolved with a sampled version of $h_m(t)$, for $0 \leq m \leq T-1$. The time resolution attained in this step is $1/B_{\text{tot}}$. 
  \item \textbf{Beamforming:} Correlations between the observation vectors from the previous step and steering vectors corresponding to each azimuth on the grid defined by the array aperture are computed. The spatial resolution attained in this step is $2/TR$.
  \item \textbf{Doppler detection:} Correlations between the resulting vectors and Doppler vectors, with Doppler frequencies lying on the grid defined by the number of pulses, are computed. The Doppler resolution is $1/P\tau$.
  \item \textbf{Peak detection:} Similar to classic radar, but detection is performed on the 3D range-azimuth-Doppler map. 
\end{enumerate}

\end{multicols*}
\end{minipage}}\end{figure*}

\subsection{MIMO pulse-Doppler Radar}

MIMO radar presents significant potential for advancing state-of-the-art modern radar in terms of flexibility and performance. This configuration \cite{fishler2004mimo} combines several antenna elements both at the transmitter and receiver. Unlike phased-array systems, each transmitter radiates a different waveform, which offers more degrees of freedom \cite{li2009mimo}. There are two main configurations of MIMO radar, depending on the location of the transmitting and receiving elements; collocated MIMO \cite{li2007mimo} in which the elements are close to each other relatively to the signal wavelength, and multistatic MIMO \cite{haimovich2008mimo} where they are widely separated. In this review, we focus on collocated pulse-Doppler MIMO systems.

Collocated MIMO radar systems exploit waveform diversity, based on mutual orthogonality of the transmitted signals \cite{li2009mimo}. Consequently, the performance of MIMO systems can be characterized by a virtual array constructed by the convolution of the locations of the transmit and receive antenna locations. In principle, with the same number of antenna elements, this virtual array may be much larger than the array of an equivalent traditional system \cite{bliss2003multiple, rabideau2003ubiquitous, li2009diversity}. 

\begin{figure}
\begin{center}
\includegraphics[width=0.5\textwidth]{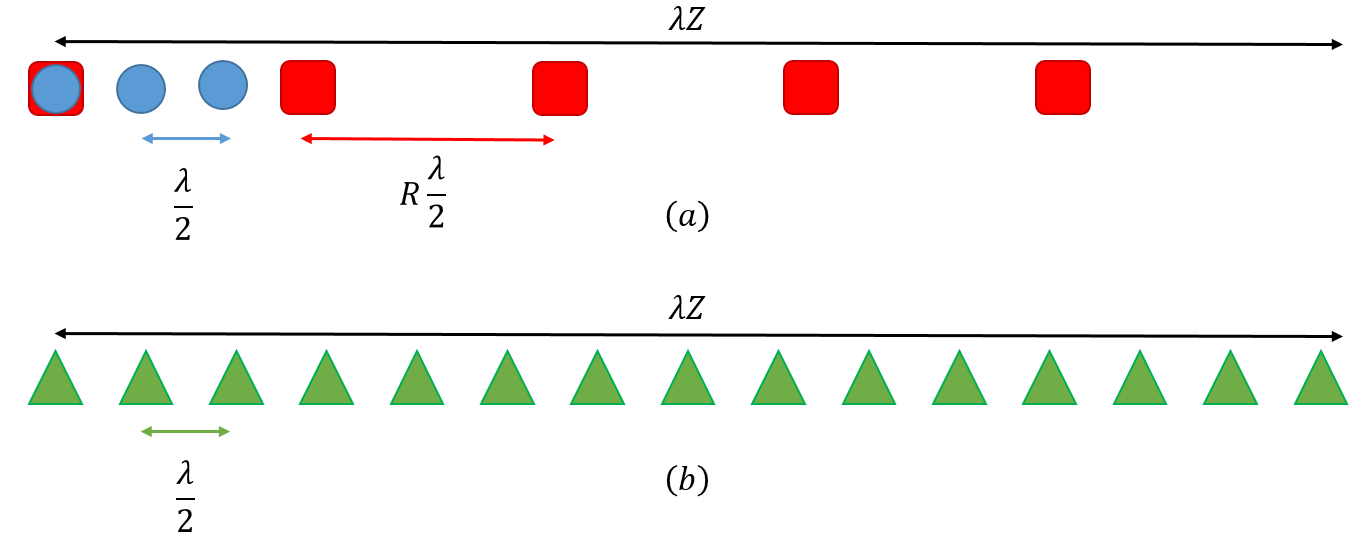}
\caption{Illustration of MIMO arrays: (a) standard array, (b) corresponding receiver virtual array \cite{cohen2016summer}.}
\label{fig:arrays1}
\end{center}
\end{figure}

The standard approach to collocated MIMO adopts a virtual uniform linear array (ULA) structure \cite{chen2009signal}, where $R$ receivers, spaced by $\frac{\lambda }{2}$ and $T$ transmitters, spaced by $R\frac{\lambda }{2}$ (or vice versa), form two ULAs. Here, $\lambda$ is the signal wavelength. Coherent processing of the resulting $TR$ channels generates a virtual array equivalent to a phased array with $TR$ $\frac{\lambda }{2}$-spaced receivers and normalized aperture $Z=\frac{TR}{2}$. Denote by $\{\xi_{m}\}_{m=0}^{T-1}$ and $\{\zeta _{q}\}_{q=0}^{R-1}$ the normalized transmitters and receivers' locations, respectively. For the traditional virtual ULA structure, $\zeta_q=\frac{q}{2}$ and $\xi_m=R\frac{m}{2}$. 
%Several approaches consider random array configurations \cite{rossi2014spatial, cohen2016summer}, where the antennas' locations are chosen uniformly at random within the aperture of the virtual array described above, that is $\{\xi_{m}\}_{m=0}^{M-1} \sim \mathcal{U} \left[ 0, Z \right]$ and $\{\zeta _{q}\}_{q=0}^{Q-1} \sim \mathcal{U} \left[ 0,Z \right]$, respectively.
This standard array structure and the corresponding virtual array are illustrated in Fig.~\ref{fig:arrays1} for $R=3$ and $T=5$. The circles represent the receivers and the squares are the transmitters.

%% MIMO waveforms
Each transmit antenna sends $P$ pulses, such that the $m$th transmitted signal is given by 
    \begin{equation}\label{trMth}
  s_m(t) = \sum_{p=0}^{P-1} {{{h}_{m}}\left( t -p\tau \right)}{{e}^{j2\pi {{f}_{c}}t}},\quad0\le t\le P\tau,
    \end{equation}
    where ${{h}_{m}}\left( t \right), 0\le m\le T-1 $ are orthogonal pulses with bandwidth $B_h$ and modulated with carrier frequency ${{f}_{c}}$. For convenience, it is typically assumed that $f_c \tau$ is an integer, so that the initial phase for every pulse $e^{-j2\pi f_c \tau p}$ is canceled in the modulation for $0 \leq p \leq P-1$ \cite{peebles2007radar}.

MIMO radar architectures impose several requirements on the transmitted waveform family. Besides traditional demands from radar waveforms such as low sidelobes, MIMO transmit antennas rely on orthogonal waveforms. In addition, to avoid cross talk between the $T$ signals and form $TR$ channels, the orthogonality condition should be invariant to time shifts, that is $\int_{-\infty }^{\infty }{{{s}_{i}}\left( t \right)s_{j}^{*}\left( t-\tau_0  \right)dt}=\delta \left( i-j \right),$ for $i,j \in \left[0,T-1\right]$ and for all $\tau_0$. The main waveform families typically considered are time, frequency and code division multiple access, abbreviated as TDMA, FDMA and CDMA, respectively. Time invariant orthogonality is achieved by FDMA and TDMA and approximately achieved by CDMA, as the latter involves overlapping frequency bands \cite{vaidyanathan2008mimo}.

Besides the traditional assumptions on the targets, MIMO systems present additional requirements on the radar array and waveforms with respect to the targets, as described in ``Targets' Assumptions". In the MIMO configuration, the goal is to recover the targets' azimuth angles $\theta_l$ in addition to their delays $\tau_l$ and Doppler shifts $\nu_l$ from the received signals. 

%%%%%%%%%%%%%%%%%%%%%%%%%%%%%%%%%%%%%%%%
% The CS BOX
%%%%%%%%%%%%%%%%%%%%%%%%%%%%%%%%%%%%%%%%
\begin{figure*}[h!]\fboxsep1em
\colorbox{BoxBackground}{\begin{minipage}{1\textwidth}\begin{multicols*}{2}
\section*{Compressed Sensing Recovery}
\label{box:cs}

CS \cite{CSBook, SamplingBook} is a framework for simultaneous sensing and compression of finite-dimensional vectors, which relies on linear dimensionality reduction. In particular, the field of CS focuses on the following recovery problem
\begin{equation}
\bf z = Ax,
\end{equation}
where $\bf x$ is a $N\times 1$ sparse vector, namely with few non zero entries, and $\bf z$ is a vector of measurements of size $M<N$. CS provides recovery conditions and algorithms to reconstruct $\bf x$ from the low-dimensional vector $\bf z$.

Two popular CS greedy recovery algorithms, orthogonal matching pursuit (OMP) and iterative hard thresholding (IHT), attempt to solve the optimization problem
\begin{equation}
\mathbf{\hat{x}}= \arg \min_\mathbf{x} ||\mathbf{x}||_0 \quad \text{s.t. } \mathbf{z = Ax},
\end{equation}
where $||\cdot ||_0$ denotes the $\ell_0$-norm. OMP \cite{pati1993orthogonal, davis1994adaptive} iteratively proceeds by finding the column of $\bf A$ most correlated to the signal residual $\bf r$,
\begin{equation}
\label{eq:omp1}
i = \arg \max |\mathbf{A}^H \mathbf{r}|,
\end{equation}
where the absolute value is computed element-wise and $(\cdot)^H$ is the Hermitian operator. The residual is obtained by subtracting the contribution of a partial estimate $\mathbf{\hat{x}}_{\ell}$ of the signal at the $\ell$th iteration, from $\bf z$, as follows:
\begin{equation}
\label{eq:omp2}
\mathbf{r}=\mathbf{z}-\mathbf{A}\mathbf{\hat{x}}_{\ell}.
\end{equation}
It is initialized by $\bf r =z$. Once the support set is updated by adding the index $i$, the coefficients of $\mathbf{\hat{x}}_{\ell}$ over the support set are updated, so as to minimize the residual error.

Other greedy techniques include thresholding algorithms. We focus here on the IHT method proposed in \cite{DaviesIHT}. Starting from an initial estimate $\mathbf{\hat{x}}_0 =0$, the algorithm iterates a gradient descent step with step size $\mu$ followed by hard thresholding, i.e.,
\begin{equation}
\label{eq:iht}
\mathbf{\hat{x}}_{\ell} = \mathcal{T}(\mathbf{\hat{x}}_{\ell-1} + \mu \mathbf{A}^H (\mathbf{z}-\mathbf{A}\mathbf{\hat{x}}_{\ell-1}), k)
\end{equation}
until a convergence criterion is met. Here $\mathcal{T}(\mathbf{x}, k)$ denotes a thresholding operator on $\bf x$ that sets all but the $k$ entries of $\bf x$ with the largest magnitudes to zero, and $k$ is the sparsity level of $\bf x$, assumed to be known.

Alternative approaches to greedy recovery are convex relaxation based methods, using $\ell_1$ regularization such as basis pursuit and LASSO. Further details on CS recovery conditions and techniques can be found in \cite{CSBook, SamplingBook}.
\end{multicols*}\end{minipage}}\end{figure*}

\subsection{Current Challenges}
Standard radar processing samples and processes the received signal at its Nyquist rate $B_h$. For example, the pulse-Doppler classic radar processing, described in ``Classic pulse-Doppler and MIMO Processing", first filters the sampled signal by a matched filter (MF). In modern systems, the MF operation is performed digitally, and therefore requires an ADC capable of sampling at rate $B_h$. Other radar systems similarly require sampling the received signal at its Nyquist rate. The radar bandwidth $B_h$ is inversely proportional to the system fast time, or range resolution, and can thus be hundreds of MHz or even up to several GHz, requiring high sampling rate and resulting in a large number of samples per pulse $N=\tau B_h$ to process.

The slow time, or Doppler, resolution is inversely proportional to the CPI $P\tau$.  The Doppler processing stage can be viewed as a MF in the pulse dimension, or slow time domain, to a constant radial velocity target. As such, it increases the signal-to-noise ratio (SNR) by $P$ compared to the SNR of a single pulse \cite{richards2005fundamentals}. Since a MF is the linear time-invariant system that maximizes SNR, it follows that a factor $P$ increase is optimal for $P$ pulses. A large number of pulses, increases resolution and SNR, but leads to large time-on-target and a large total number of samples to process given by $PN$.
 
The required computational power corresponds to $P$ convolutions of a signal of length $N=\tau B_h$ and $N$ fast Fourier transforms (FFT) of length $P$ (see ``Classic Pulse Doppler and MIMO Processing"). The growing demands for improved estimation accuracy and target separation dictate an ever growing increase in signal's bandwidth and CPI. This creates bottlenecks in sampling and processing rates in the fast time, or intra-pulse, domain, and time-on-target in the slow time, or inter-pulse, dimension.

In MIMO radar, the additional spatial dimension increases the system's complexity, as may be seen in ``Classic pulse-Doppler and MIMO Processing". In such systems, the array aperture determines the azimuth resolution. In a traditional virtual array configuration, the product between the number of transmit and receive antennas scales linearly with the aperture. Consequently, high resolution requires a large number of antennas, increasing the system's complexity in terms of hardware and processing.

In the following sections, we review fast time compressed radar systems that allow for low rate sampling and processing of radar signals, regardless of their bandwidth, while retaining the same SNR scaling. We then demonstrate how compression can be extended to the slow time, reducing time-on-target, and to the spatial dimension allowing one to achieve similar resolution as a filled array, but with significantly fewer elements. In reality, the received signal $x_R(t)$ is further contaminated by additive noise and clutter. We will thus also demonstrate the impact of SNR and clutter on compressed radar system prototypes \cite{eldar2015clutter, barilan2014focusing, baransky2014prototype}. Finally, we show how compression and sub-Nyquist sampling can be exploited to address other challenges, such as communication and radar spectrum sharing.

\section{Increased Parameter Resolution}

In many radar applications, the reflectivity scene consists of a small number $L$ of strong targets. Therefore, CS techniques (see ``Compressed Sensing Recovery") are a natural processing tool for radar systems. Shortly after the idea of CS was brought forward by the works of Candes, Romberg and Tao \cite{candes2006robust, candes2006stable}, and of Donoho \cite{donoho2006compressed} a decade ago, it was introduced to pulse-Doppler radar \cite{herman2009high, chi2009golay, baraniuk2007compressive} and SFR \cite{shah2009step}.

While CS is typically applied to signal processing tasks to reduce the associated sampling rate~\cite{SamplingBook}, earlier papers that applied CS recovery to pulse-Doppler radar and SFR were aimed at increasing delay-Doppler resolution \cite{herman2009high, chi2009golay, shah2009step, gedalyahu2011identification} using Nyquist samples. More recent approaches use CS recovery techniques on low rate, or sub-Nyquist samples, enabling sampling and processing rate reduction while achieving the same resolution as traditional Nyquist radars. In the remainder of this section, we review radar recovery methods that increase delay-Doppler resolution using CS techniques on Nyquist samples. In the next sections, we consider the application of CS to reduce the fast time sampling rate, number of pulses and antennas while preserving the resolution achieved by Nyquist systems.

In the works of \cite{herman2009high, chi2009golay, shah2009step, gedalyahu2011identification}, the signal is still sampled at its Nyquist rate $B_h$ but the delay and Doppler resolutions are determined by the CS grid, containing $N>\tau B_h$ grid points, rather than the signal's bandwidth and CPI, respectively. The key idea in \cite{herman2009high}, which adopts a pulse-Doppler radar model, is that the received signal $x_R(t)$ defined in (\ref{eq:uni_rec}) is generally a sparse superposition of time-shifted and frequency-shifted replicas of the transmitted waveforms. The time-frequency plane is discretized into a $N \times N$ grid where each point represents a unique time-frequency shift $\mathbf{H}_i$, expressed as the product of time-shift and frequency modulation matrices, denoted by $\mathbf{T}^{(.)}$ and $\mathbf{M}^{(.)}$, respectively. In particular,
\begin{equation}
\mathbf{H}_i=\mathbf{M}^{i\text{mod}N} \mathbf{T}^{\lfloor i/N \rfloor},
\end{equation}
where
\begin{equation} \label{eq:matTandM}
\mathbf{T} =\begin{pmatrix}
   0 & 0 &  & 1 \\
1 & 0 & & 0 \\
& \ddots & \ddots & \\
 0 & & 1 & 0
  \end{pmatrix}, \quad 
\mathbf{M} =\begin{pmatrix}
   1 &  &  & 0 \\
 & e^{j \frac{2 \pi}{N}} & &  \\
& & \ddots & \\
 0 & &  & e^{j \frac{2 \pi}{N} (N-1)}
\end{pmatrix}.
\end{equation}
Here, $\lfloor \cdot \rfloor$ and $\text{mod}$ denote the floor and modulation functions, respectively.

The vector $\bf y$, that concatenates the Nyquist samples of a single pulse $x_p(t)$, can then be expressed as
\begin{equation} \label{eq:Nyquist_samples}
\bf y = \Phi s,
\end{equation}
where $\bf s$ is the $L$-sparse vector of size $N^2$ whose non zero entries are the targets' RCS $\alpha_l$ with locations determined by the corresponding time-frequency shift. The $i$th column, or atom, of the $N \times N^2$ matrix $\bf \Phi$ is given by
\begin{equation}
\mathbf{\Phi}_i=\mathbf{H}_i \mathbf{f},
\end{equation} 
where the vector $\bf f$ contains the Nyquist rate samples $h[n]$ of the transmitted signal $h(t)$. The latter is chosen so that the samples correspond to the Alltop sequence $h[n]=\frac{1}{\sqrt{N}} e^{2 \pi j n^3/N}$ \cite{alltop1980complex}, for some prime $N \geq 5$. This yields a low coherence matrix $\bf \Phi$, namely a matrix whose columns have small correlation. 

The vector $\bf s$ is reconstructed from $\bf y$ using CS techniques, as described in ``Compressed Sensing Recovery". The time-frequency shifts, determined by the targets' delays and Doppler frequencies, are thus recovered with a resolution of $1/N$.

The CS recovery in \cite{herman2009high} is performed without a MF, which reduces performance in low SNR regimes. In addition, \cite{herman2009high} considers only delay recovery. Alternatively, CS techniques can be performed after applying an analog MF \cite{chi2009golay} on the pulse-Doppler received signal (\ref{eq:uni_rec}). The MF output of the $p$th pulse, sampled at the Nyquist rate $1/B_h$, is given by
\begin{equation}
w_p[k]= \sum_{l=0}^{L-1} \alpha_l e^{j \nu_l \tau_l} e^{j \nu_l p \tau} C_h[k-\tau_l/\tau],
\end{equation}
where $C_h[k]$ is the discrete autocorrelation function of the transmitted waveform. For each sampling time $k$, the Nyquist samples have a sparse representation in the frequency, or Doppler, domain using a Fourier matrix as dictionary. A two-step approach is thus proposed that applies CS recovery for each $k$. However, the sidelobes of $C_h[k]$ lead to ambiguity. To avoid these, pairs of Golay complementary sequences $x_1$ and $x_2$ of length $N$, whose correlation functions satisfy
\begin{equation}
C_{x_1}[k]+C_{x_2}[k]=2N\delta[k],
\end{equation}
are transmitted alternatively, by phased-coding the baseband waveform. This allows for unambiguous delay-Doppler recovery provided that all Doppler coordinates are within the interval $[-\pi/2, \pi/2]$.

CS has also been applied to SFR in order to increase range resolution \cite{shah2009step}. As in pulse-Doppler radar, the target scene is discretized over a $N \times N$ delay-Doppler map \cite{shah2009step}. The outputs of the phase detector (\ref{eq:sfr_phase_detector}) are then expressed as in (\ref{eq:Nyquist_samples}) where $\bf y$ is the vector of size $P$ with the $p$th entry given by $y_p$ and $\bf \Phi$ is a DFT based dictionary such that
\begin{equation}
\mathbf{\Phi}_{(p,(i-1)N+k)}=e^{j 2 \pi f_p \tau_i}e^{j \nu_k p \tau}.
\end{equation}
The vector $\bf s$ is then recovered from $\bf y$ using CS techniques.

The approaches above may increase resolution by taking a large grid size $N$. However, bounds on $N$ are not discussed and it is not clear how large it can be. Denser grids reduce sensitivity of the reconstruction to off-grid targets but increase the computational complexity by a square factor since the dictionaries contain $N^2$ atoms. More importantly, higher grid dimensions cause significant increase to the coherence of the CS dictionary, which may degrade recovery performance.

The parameter space discretization, typically involved in CS recovery techniques, assumes the targets' delays and Dopplers lie on the predefined grid. Several approaches have been proposed to solve off-grid issues. These include grid refinement that adjusts the detected delay-Doppler peak \cite{cohen2016summer}, parameter perturbation based adaptive sparse reconstruction techniques \cite{teke2014robust}, and sensing matrix perturbation \cite{yang2012robustly}. More references may be found in \cite{hadi2015compressive}.

\section{Fast Time Compression}

In the works we reviewed so far, sampling and digital processing are still performed at the Nyquist rate. We next consider compressed radar that reduces sampling and processing rates.

\subsection{Random Sampling}

Random sampling has been considered in SFR systems by selecting random measurements out of the Nyquist samples \cite{teke2014robust, ender2010compressive}. The SFR approach of (\ref{eq:sfr}) is adopted in \cite{ender2010compressive}, with a random selection of $M$ out of $P$ pulses with different carriers. The sparse representation of the received signal used is a delay-Doppler shifted dictionary \cite{teke2014robust} similar to \cite{herman2009high}. Consider the matrix $\bf \Phi$ whose $i$th column is given by
\begin{equation}
\label{eq:dico_random}
\mathbf{\Phi}_i = h(\mathbf{t}-\tau_i) \circ e^{j2\pi \nu_i \mathbf{t}},
\end{equation}
where $\bf t$ is the $N \times 1$ vector containing the sampling instants at the Nyquist rate, that is $t_i=i/B_h$, and $\circ$ is the Hadamard product operator. As in \cite{herman2009high}, the dictionary $\bf \Phi$ contains $N^2$ atoms.
The Nyquist samples can then be expressed in the form (\ref{eq:Nyquist_samples}) and the compressed samples $\bf z$ are given by
\begin{equation}
\bf z=Ay,
\end{equation}
where $\bf A$ is a $M \times N$ matrix, with $M < N$, constructed by randomly selecting $M$ rows of the $N \times N$ identity matrix, which corresponds to the $M$ selected pulses.

In these approaches, processing is performed at a low rate. However, the random discarding of samples is difficult to implement in a sampling system to effectively reduce the sampling rate. Furthermore, the large dictionary size discussed in the previous section remains an issue. Alternative practical radar systems using CS to reduce the sampling rate have been proposed, that rely on two main techniques: uniform low rate sampling using appropriate waveforms and analog pre-processing.

\subsection{Uniform Low Rate Sampling}

In \cite{ertin2011sparse}, the authors consider SFR using multi-frequency chirps, described in (\ref{eq:trans_chirp}). Low rate samples are uniformly taken from the received signal (\ref{eq:chirp_sig}) at rate $2\gamma \tau_r$, with $\tau_r=2 (r_{\text{max}}-r_{\text{min}})/c$, and $\gamma$ being the common chirp-rate. This results in aliasing of the multiple sinusoids to baseband, with random complex coefficients. Upon discretization of the target range, denoted by $\bf s$, the low rate samples may be modeled as
\begin{equation}
\mathbf{y=As}.
\end{equation}
Here, the $k$th column of the sensing matrix $\bf A$ is the FFT of the samples of (\ref{eq:chirp_sig}) for a singular target at range bin $k$ corresponding to a delay of $\tau_l=2(r_{\text{min}}-k\Delta)/c$, where $\Delta$ is the range discretization step. The targets' delays are thus recovered from low rate uniform sampling of the chirp waveforms.

\subsection{Random Demodulation}

Many analog-to-information conversion (AIC) systems have been proposed to sample wideband signals at sub-Nyquist rates. Among them, the random demodulator (RD)  \cite{RandomDemodulator2}, random pre-integrator (RMPI) \cite{yoo2012design} and Xampling-based \cite{mishali2011xampling, Xampling} systems have been used for radar applications. All three approaches consider pulse-Doppler radar.

\begin{figure}
\begin{center}
\includegraphics[width=1\columnwidth]{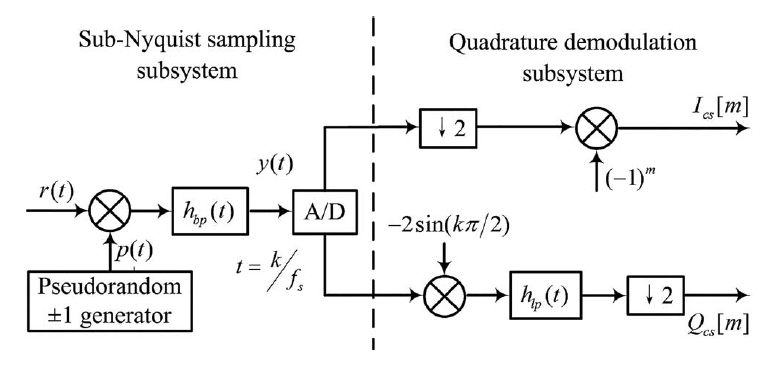}
\caption{QuadCS implementation with RD sampling (left pane) followed by quadrature demodulation (right pane) \cite{liu2015pulse}.}
\label{fig:randomD}
\end{center}
\end{figure}

The RD modulates the input signal using a high-rate sequence $p(t)$ created by a pseudo-random number generator, aliasing its frequency content. The random sequence used for demodulation is a square wave, which alternates between the levels $\pm 1$ with equal probability. The mixed output is filtered by a bandpass filter $h_{bp}(t)$, with center frequency $f_c$ and bandwidth $B_{CS} \ll B_h$, and sampled at a low rate, as shown in the left pane of Fig.~\ref{fig:randomD}.

The RD is adopted in \cite{liu2015pulse} as the analog mixing front-end of a proposed quadrature compressive sampling (QuadCS) system. The mixed and filtered output $y(t)$, shown in Fig.~\ref{fig:randomD}, is given by
\begin{equation}
y(t)=\int_{-\infty}^{\infty} h_{bp}(\rho) p(t-\rho) x_R(t-\rho) \mathrm{d}\rho,
\end{equation}
where $x_R(t)$ is defined as the real part of (\ref{eq:uni_rec}). The RD samples $y(t)$ at rate $f_s=1/T_s=f_c/k$ with $k$ an integer satisfying $k \leq \lfloor f_c/2B_{CS} \rfloor$. The samples are fed to the quadrature processing system~\cite{liu1989new}, which extracts the baseband in-phase and quadrature (I and Q) components of the radar echoes. As shown in~\cite{liu2015pulse}, the complex samples of the RD output can be written as
\begin{equation}
\mathbf{y}=\mathbf{A} \mathbf{x}.
\end{equation}
Here, $\bf x$ is a sparse vector that contains the complex amplitudes $\alpha_l$ at the corresponding delays $\tau_l$ and the $(m,p)$ element of the matrix $\bf A$ is given by
\begin{equation}
\label{eq:dico_RD}
\mathbf{A}_{m,p}= \int_{-\infty}^{\infty} h_{bp}(\rho) e^{-j2\pi f_c\rho} p(mT_s - \rho) h(mT_s - p\tau - \rho) \mathrm{d}\rho.
\end{equation}
The samples of $P$ pulses are concatenated in a matrix $\mathbf{Y}$ such that each column corresponds to a pulse. The subsequent processing of the QuadCS, referred to as compressive sampling pulse-Doppler (CoSaPD), is composed of a DFT step on the rows of $\bf Y$ that acts as a MF in slow time followed by a MF in each column, corresponding to the fast time.

The random-modulation pre-integrator (RMPI) is a variant of the RD composed of a parallel set of RD channels driven by a common input, where each RD uses a distinct pseudo-random binary sequence (PRBS). A hardware RMPI-based prototype has been implemented in~\cite{yoo2012compressed}, which recovers radar pulses and estimates their amplitude, phase and carrier frequency. In the next section, we show an alternative prototype with a different analog front end, which also recovers the targets' parameters from low rate samples.

It is interesting to note that the considerations behind waveform design for CS recovery in the approaches~\cite{herman2009high, chi2009golay, shah2009step} presented in the previous section are similar to traditional radar requirements. The well known ambiguity function (AF) impacts CS radar in a similar way as traditional radar systems. Indeed, the mutual coherence of the dictionary is linearly related to the highest side lobe value of the AF \cite{hadi2015compressive, song2010role}. In contrast, we will see in the next section that the CS dictionary of the Xampling method is independent of the waveform, and MF is performed directly on the low rate samples before parameter recovery.

%%%%%%%%%%%%%%%%%%%%%%%%%%%%%%%%%%%%%%%%
% The Doppler Focusing BOX
%%%%%%%%%%%%%%%%%%%%%%%%%%%%%%%%%%%%%%%%
\begin{figure*}[h!]\fboxsep1em
\colorbox{BoxBackground}{\begin{minipage}{1\textwidth}\begin{multicols*}{2}
\section*{Doppler Focusing}
\label{box:cs}

Doppler focusing is a processing technique, suggested in \cite{barilan2014focusing}, which uses target echoes from different pulses to create a superimposed pulse focused at a particular Doppler frequency. This method allows for joint delay-Doppler recovery of all targets present in the illuminated scene. It results in an optimal SNR boost, and may be carried out in the frequency domain, thus enabling sub-Nyquist sampling and processing with the same SNR increase as a MF.

The output of Doppler processing can be viewed as a discrete equivalent of the following time shift and modulation operation on the received signal:
\begin{equation}
\label{eq:doppler_focus}
\Phi(t,\nu)=\sum_{p=0}^{P-1} x_p(t + p\tau) e^{j \nu p \tau}
=  \sum_{l=0}^{L-1} \alpha_l h(t-\tau_l) \sum_{p=0}^{P-1} e^{j (\nu-\nu_l) p \tau}.
\end{equation}
Consider the sum $g(\nu|\nu_l) = |\sum_{p=0}^{P-1} e^{j (\nu-\nu_l) p \tau}|$.
For any given $\nu$, targets with Doppler frequencies $\nu_l$ in a band of width $2\pi/P \tau$ around $\nu$ will achieve coherent integration and an SNR increase of approximatively $P$. On the other hand, since the sum of $P$ equally spaced points covering the unit circle is generally close to zero, targets with $\nu_l$ not ``in focus" will approximately cancel out. In summary, we have that
\begin{equation} \label{eq:focus_approx}
g(\nu|\nu_l) = \sum_{p=0}^{P-1} e^{j (\nu-\nu_l) p \tau} \approx \left\{ \begin{array}{ll} 
P & |\nu -\nu_l| < \pi /P \tau \\
0 & |\nu -\nu_l| \geq \pi /P \tau,
\end{array} \right.
\end{equation}
as shown in Fig.~\ref{fig:sum_exp1}.
\begin{center}
\includegraphics[width=0.35\textwidth]{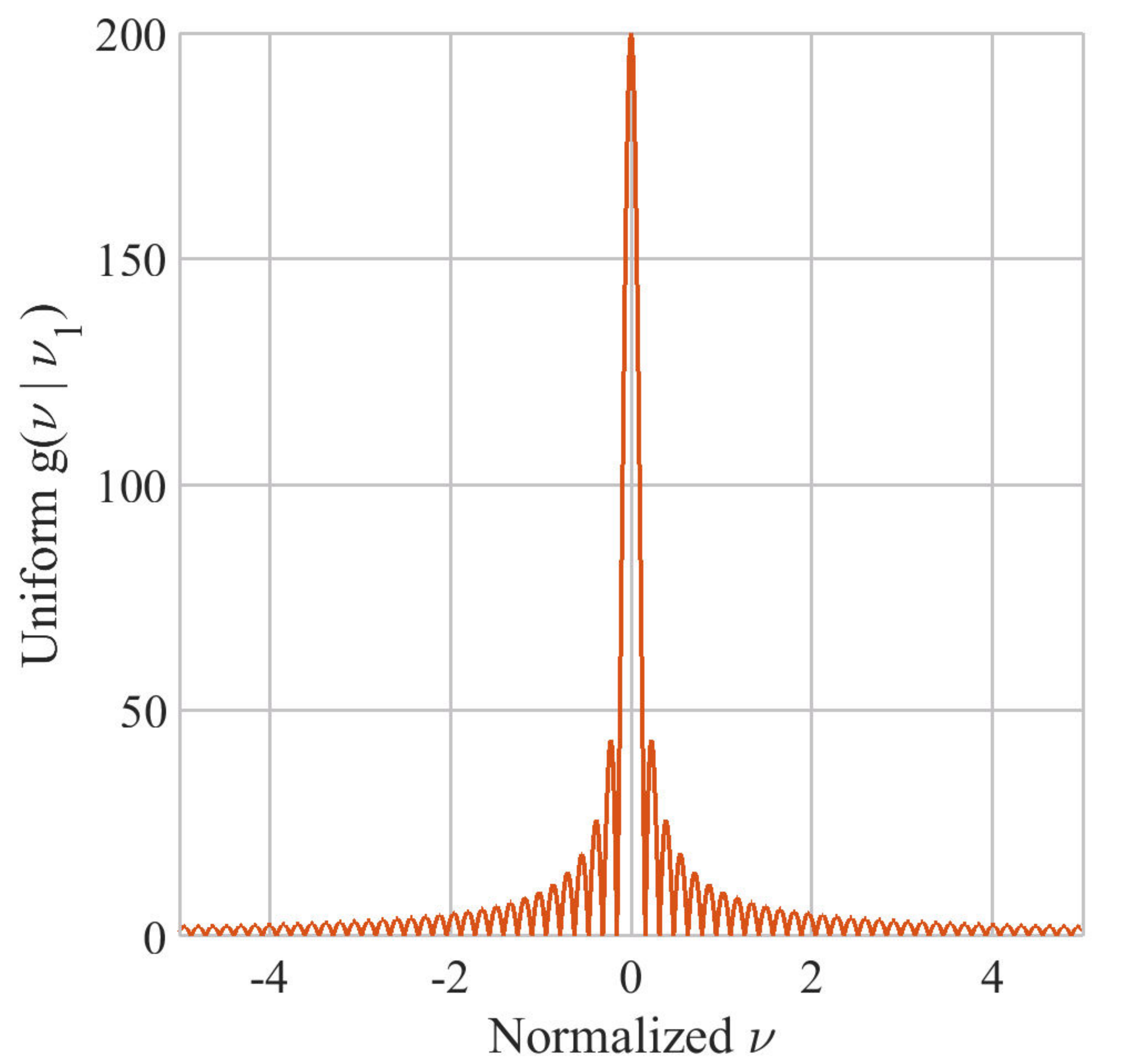}
\caption{Sum of exponents $|g(\nu|\nu_l)|$ for $P=200$, $\tau=1$sec and $\nu_l=0$.}
\label{fig:sum_exp1}
\end{center}
We may therefore estimate the sum of exponents in (\ref{eq:doppler_focus}) as
\begin{equation}
\Phi(t,\nu) \approx P \sum_{l \in \Lambda(\nu)} \alpha_l h(t-\tau_l),
\end{equation}
where $\Lambda(\nu) = \{l: |\nu-\nu_l| < \pi/ P \tau \}$. In other words, the sum is only over the targets whose Doppler shifts are in the interval $|\nu-\nu_l| < \pi/ P \tau$.

For each Doppler frequency $\nu$, $\Phi(t, \nu)$ represents a standard pulse-stream model where the problem is to estimate the unknown delays. Thus, using Doppler focusing, the two-dimension delay-Doppler recovery problem is reduced to delay-only estimation for a small range of Doppler frequencies, with increased SNR by a factor of $P$ \cite{SamplingBook}. 
The Xampling-radar of \cite{barilan2014focusing} performs Doppler focusing directly on the low rate samples in the frequency domain, allowing for joint Doppler-delay recovery from the ``Xamples". 
\end{multicols*}\end{minipage}}\end{figure*}

\subsection{Fast Time Xampling}
An alternative sub-Nyquist radar method is the Xampling-based system proposed in~\cite{barilan2014focusing, baransky2014prototype}. As we show in this section, this approach, which may be used with any transmitted pulse shape, achieves the minimal sampling rate required for target detection while providing optimal SNR. 

The sub-Nyquist analog front-end is composed of an ADC which filters the received pulse-Doppler signal (\ref{eq:uni_rec}) to predetermined frequencies before taking point-wise samples. These compressed samples, or ``Xamples", contain the information needed to recover the desired signal parameters, that is the targets' delay-Doppler map. To see this, note that the Fourier series coefficients of the aligned frames $x_p(t+p \tau)$ are given by
\begin{equation}
\label{eq:fourier_coeff}
c_p[k] = \frac{1}{\tau} H[k] \sum_{l=0}^{L-1} \alpha_l e^{-j 2 \pi k \tau_l / \tau} e^{-j \nu_l p \tau}, \quad 0 \leq k \leq N-1,
\end{equation}
where $H[k]$ are the Fourier coefficients of the known transmitted pulse $h(t)$ and $N=B_h\tau$ is the number of Fourier samples. From (\ref{eq:fourier_coeff}), we see that the unknown parameters $\{ \alpha_l, \tau_l, \nu_l \}_{l=0}^{L-1}$ are contained in the Fourier coefficients $c_p[k]$. We now show how the Fourier coefficients $c_p[k]$ may be obtained from low-rate samples of $x_p(t)$ and how the targets' parameters can then be recovered from $c_p[k]$ (more details may be found in \cite{barilan2014focusing}).

\begin{figure}
\begin{center}
\includegraphics[width=1\columnwidth]{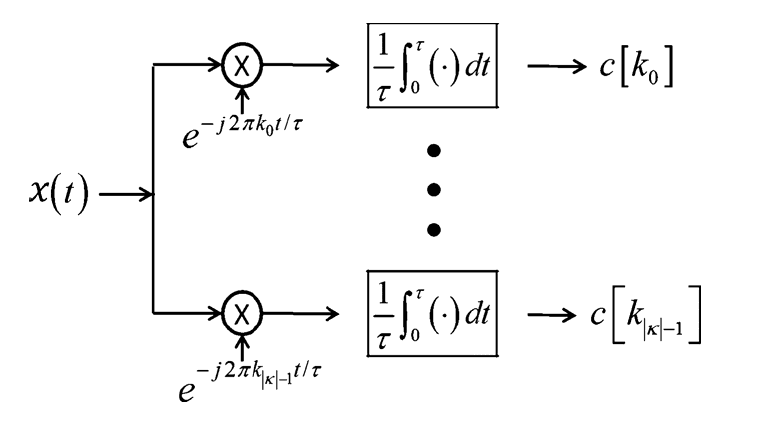}
\caption{Multichannel direct sampling of the Fourier coefficients \cite{gedalyahu2011multichannel}.}
\label{fig:multichannel}
\end{center}
\end{figure}

The received signals $x_p(t)$ exist in the time domain, and therefore there is no direct access to $c_p[k]$. To obtain any arbitrary set of Fourier series coefficients, the direct multi-channel sampling scheme \cite{gedalyahu2011multichannel}, illustrated in Fig.~\ref{fig:multichannel}, can be used. The analog input $x_p(t)$ is split into $K=|\kappa|$ channels, where in each channel $k_i$ with $i \in [0,K-1]$, it is mixed with the harmonic signal $e^{-j 2\pi k_it/ \tau}$, integrated over the PRI duration and then sampled. Xampling thus allows one to obtain an arbitrary set $\kappa$ out of $N=\tau B_h$ frequency components from $K$ point-wise samples of the received signal after appropriate analog preprocessing. An alternative Xampling method uses the Sum of Sincs filter described in \cite{tur2011innovation}. This class of filters, which consists of a sum of sinc functions in the frequency domain, is a general sampling scheme for arbitrary pulse shapes.

A less expensive and more practical approach for the Fourier series coefficients acquisition, proposed in \cite{baransky2014prototype}, is based on multiple bandpass filters and is adopted in the Xampling hardware radar prototype, described in the next section. Briefly, this system is composed of a few channels, each sampling the content of a narrow frequency band of the received signal. Each channel thus yields a group of several consecutive Fourier coefficients.
The multiple bandpass constellation has the advantage of acquiring the measurements over a wider frequency aperture. At the same time, it still allows practical hardware implementation, as detailed in the next section. By widening the frequency aperture, a finer resolution grid may be employed during the recovery process. Moreover, empirical results show that highly distributed frequency samples provide better noise robustness \cite{baransky2014prototype}. However, widening the frequency aperture eventually requires increasing the number of samples $K$, otherwise recovery performance may degrade. This trade-off is observed in the experiments presented in \cite{baransky2014prototype}. 

Once a set of Fourier coefficients $c_p[k]$ has been acquired, the delays and Doppler frequencies can be recovered using different techniques. Doppler focusing \cite{barilan2014focusing}, summarized in ``Doppler Focusing", is one approach, that has several advantages, as detailed below. This method uses target echoes from all pulses to generate a focused pulse at a specific Doppler frequency. It then jointly recovers the delay-Doppler map by reducing the detection problem to a one-dimensional delay-only estimation. Performing the Doppler focusing operation in frequency results in computing the DFT of the coefficients $c_p[k]$ in the slow time domain:
\begin{eqnarray}
\label{eq:focused_coeff}
\Psi_{\nu}[k] &=&\sum_{p=0}^{P-1} c_p[k] e^{j \nu p \tau} \nonumber \\
&=& \frac{1}{\tau} H[k] \sum_{l=0}^{L-1} \alpha_l e^{-j 2 \pi k \tau_l / \tau} \sum_{p=0}^{P-1} e^{j (\nu-\nu_l) p \tau}.
\end{eqnarray}
Note that $\Psi_{\nu}[k]$ is the Fourier series of $\Phi(t, \nu)$, defined in (\ref{eq:doppler_focus}), with respect to $t$. Following the same argument as in  (\ref{eq:focus_approx}),  we have
\begin{equation} \label{eq:focus_delay}
\Psi_{\nu}[k] \approx \frac{P}{\tau}  H[k]  \sum_{l \in \Lambda(\nu)} \alpha_l e^{-j 2 \pi k \tau_l / \tau}.
\end{equation}

The resulting equation (\ref{eq:focus_delay}) is a standard delay estimation problem for each $\nu$ and may be solved using multiple techniques (see \cite{SamplingBook} for more details). However, improved performance can be obtained by jointly processing the sequences $\{\Phi_{\nu}[k]\}$ for different values of $\nu$. Thus, instead of searching separately for each of the delays $\tau_l, l \in \Lambda (\nu)$, the $L$ delays are estimated by joint processing over all Doppler frequencies.

A particularly convenient method in this case is to employ a matching pursuit type approach where the strongest peak over all $\nu$, assuming a single delay, is first found:
\begin{equation}
(\hat{\tau}_l, \hat{\nu}_l) = \arg \max_{\tau_l, \nu_l} \left| \sum_{k \in \kappa} \Psi_{\nu_l}[k] e^{j 2 \pi k \tau_l/ \tau} \right|.
\end{equation}
Once the optimal values $\hat{\tau}_l$ and $\hat{\nu}_l$ are determined, their influence is subtracted from the focused sub-Nyquist samples as
\begin{equation}
\Psi'_{\nu}[k]= \Psi_{\nu}[k] - \frac{1}{\tau} \hat{\alpha}_l e^{-j 2 \pi k \hat{\tau}_l/\tau} \sum_{p=0}^{P-1} e^{j (\nu-\hat{\nu}_l) p \tau},
\end{equation}
where 
\begin{equation}
\hat{\alpha}_l = \frac{\tau}{P |\kappa|} \sum_{k \in \kappa} \Psi_{\hat{\nu}_l}[k] e^{j 2 \pi k \hat{\tau}_l/ \tau}.
\end{equation}
The same operations are performed iteratively to find all the desired $L$ peaks. This approach does not require discretization of the targets' parameters and these are recovered over the continuous domain from a minimal number of samples.

In practice, the search for peaks can be limited to a grid, which allows to carry out all computations using simple FFT operations. Suppose we limit ourselves to the Nyquist grid, namely the grid defined by the Nyquist resolution so that $\tau_l/\tau=s_l/N$, where $s_l$ is an integer satisfying $0 \leq s_l \leq N-1$. Then, (\ref{eq:focused_coeff}) is approximately written in vector form as
\begin{equation}
\label{eq:doppler_foc}
\mathbf{\Psi}_{\nu} = P \mathbf{H} \mathbf{F}_N^K \mathbf{a}_\nu,
\end{equation}
where $\mathbf{\Psi}_{\nu} = \left[ \Psi_{\nu}[k_0] \dots \Psi_{\nu}[k_{K-1}] \right], k_i \in \kappa$ for $0 \leq i \leq K-1$, $\bf H$ is a diagonal matrix that contains the Fourier coefficients $H[k]$ of the transmitted waveforms and $\mathbf{F}_N^K$ is the partial Fourier matrix that contains the $K$ rows of the $N \times N$ Fourier matrix indexed by $\kappa$. The entries of the $L$-sparse vector $\mathbf{a}_{\nu}$ are the values $\alpha_l$ at the indices $s_l$ for the Doppler frequencies $\nu_l$ in the ``focus zone", that is $|\nu -\nu_l| < \pi / P \tau$. The $P$ equations (\ref{eq:doppler_foc}) are simultaneously solved using CS based algorithms, where in each iteration, the maximal projection of the observation vectors onto the measurement matrix is retained. More details are given in \cite{barilan2014focusing}.

\begin{figure}
\begin{center}
\includegraphics[width=1\columnwidth]{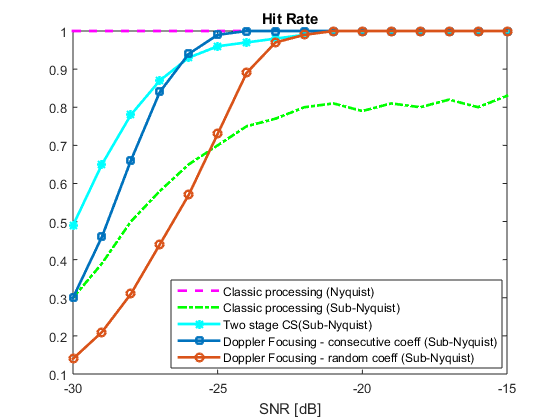}
\caption{Hit rate for classic processing, two-stage CS recovery and Doppler focusing for a fixed false alarm rate. A ``hit" is defined as a delay-Doppler estimate circumscribed by an ellipse around the true target position in the time-frequency plane, with axes equivalent to $\pm 3$ times the time and frequency Nyquist bins. The two-stage CS recovery separates delay and Doppler estimation, performing them in parallel (see~\cite{barilan2014focusing} for more details). The sub-Nyquist sampling rate was one tenth of the Nyquist rate~\cite{barilan2014focusing}.}
\label{fig:res_fast}
\end{center}
\end{figure}

Some results comparing different configurations of low rate sampling and processing are shown in Fig.~\ref{fig:res_fast} \cite{barilan2014focusing}. The recovery performance of the classic processing applied to Nyquist samples is presented as a baseline. Sub-Nyquist approaches, performed at one tenth of the Nyquist rate, include the same classic processing applied to sub-Nyquist samples, a two-stage CS recovery method that performs delay and Doppler estimation in parallel, separately (see \cite{barilan2014focusing} for details) and Doppler focusing. It is clearly seen that Doppler focusing applied to random Fourier coefficients, which are with high probability widely distributed, leading to a wide aperture, outperforms other sub-Nyquist approaches. The use of consecutive coefficients yields small aperture and poor resolution.

The Xampling approach has several advantages. First, it recovers the targets' parameters directly from the low rate samples, without requiring sampling at the Nyquist rate. Second, previous CS-based methods typically impose constraints on the radar transmitter, which are not needed here. Indeed, as may be seen in (\ref{eq:dico_random}) and (\ref{eq:dico_RD}), for example, the CS dictionary depends on samples of the waveform $h(t)$, such that the mutual coherence of the dictionary is linearly related to the highest side lobe value of the AF \cite{hadi2015compressive, song2010role}. In contrast, the CS dictionary of the Xampling method is independent of the waveform, as shown in (\ref{eq:doppler_foc}). Third, in the presence of additive white noise, Doppler focusing achieves an increase in SNR by a factor of $P$ (a detailed analysis may be found in \cite{barilan2014focusing}). In addition, this approach can operate at the minimal possible sampling rate for recovering the targets' parameters, as derived in \cite{barilan2014focusing}. The minimal number of samples required for perfect recovery of $\{\alpha_l, \tau_l, \nu_l\}$ with $L$ targets in a noiseless environment is $4L^2$, with at least $K \geq 2L$ samples per pulse and at least $P \geq 2L$ pulses. The Doppler focusing approach achieves this minimal number of samples. Finally, Doppler focusing is able to deal with certain models of clutter and target dynamic range by adding a simple windowing operation in the sum (\ref{eq:focused_coeff}) and by pre-whitening in frequency \cite{eldar2015clutter}. 

The Xampling radar was implemented in hardware, as described in the next section, demonstrating real compressed radar capabilities. The hardware prototype is built from off-the-shelf components, which are bandpass filters and low rate samplers, leading to low hardware complexity.

\subsection{Hardware Prototype}

\begin{figure}
\begin{center}
\includegraphics[width=1\columnwidth]{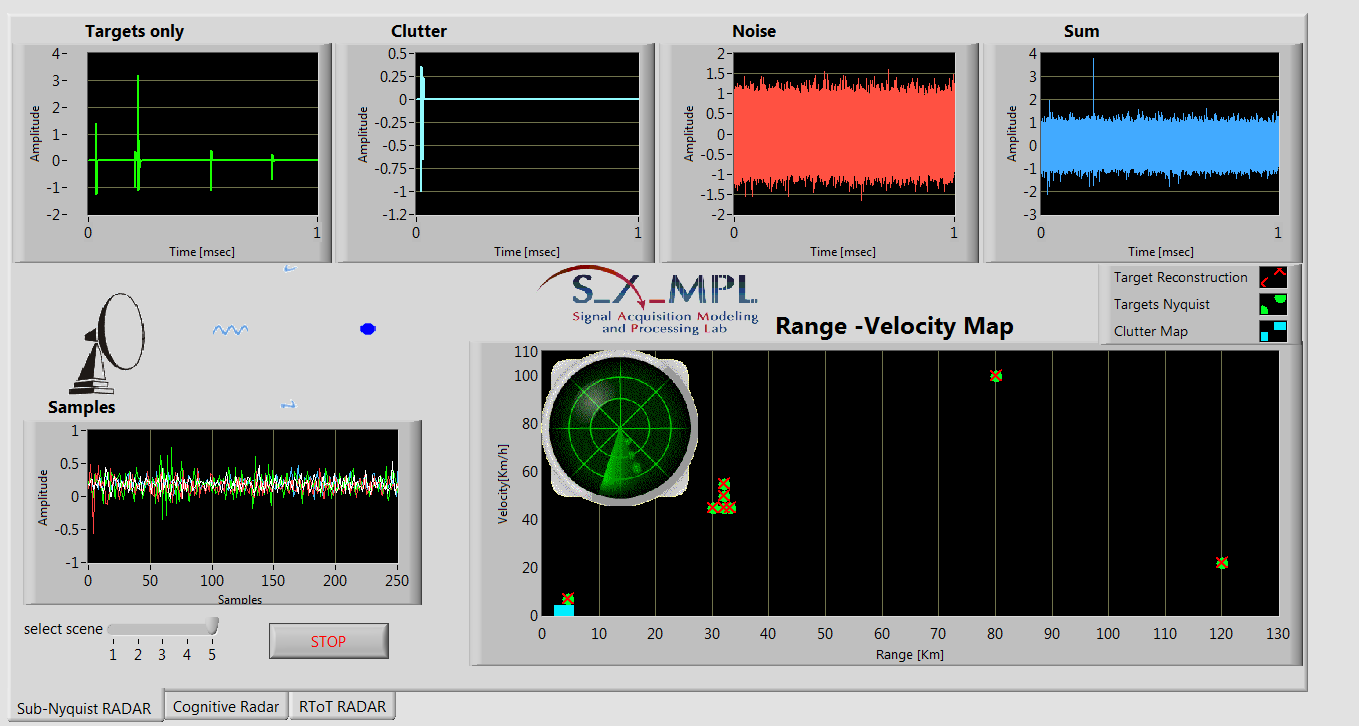}
\caption{Xampling radar LabView\textsuperscript{\textregistered} experimental interface. From left to right: at the top, received signal from targets only, received signal from clutter, noise, overall received signal $x_p(t)$. At the bottom, sub-Nyquist samples of the $4$ channels at $1/30$ of the Nyquist rate, true and recovered delay-Doppler maps. As can be seen, all targets, including close targets both in range and velocity, are correctly detected.}
\label{fig:screenshot1}
\end{center}
\end{figure}

Xampling is used in combination with Doppler focusing in the sub-Nyquist prototype of \cite{barilan2014focusing, baransky2014prototype}, which demonstrates radar reception at sub-Nyquist rates. The input signal simulates reflections from arbitrary targets and is corrupted by additive noise and clutter. The radar receiver implements the multichannel topology described in the previous section and samples a signal with Nyquist rate of $30$ MHz with a compression factor of $30$. Hardware experiments demonstrate the feasibility of detecting targets from low rate samples of an analog radar signal, using standard radio frequency (RF) hardware~\cite{barilan2014focusing, baransky2014prototype}. Typical experiment results are shown in Fig.~\ref{fig:screenshot1}, which depicts the input signal, the low rate samples and the original and recovered delay-Doppler maps, including close targets both in terms of range and velocity.

\begin{figure*}
\begin{center}
\includegraphics[width=1\textwidth]{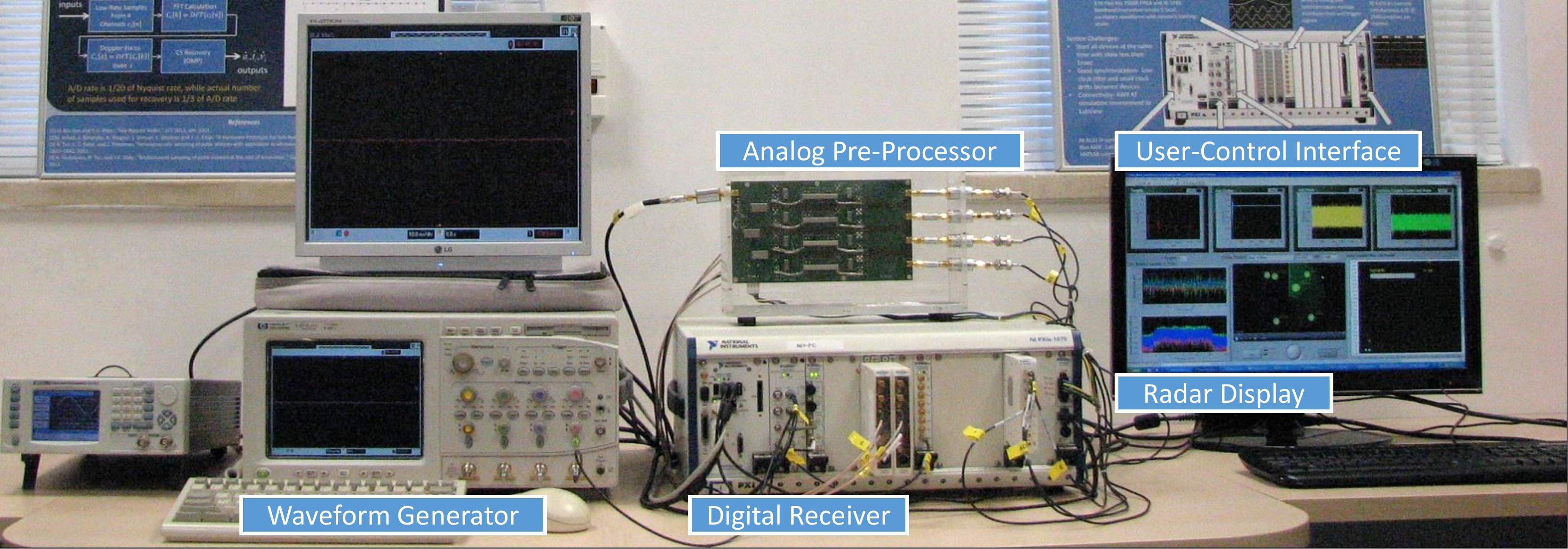}
\caption{Xampling radar prototype including AWG, receiver board, NI chassis and display \cite{baransky2014prototype}.}
\label{fig:proto_single}
\end{center}
\end{figure*}

At the heart of the receiver lies the Xampling based ADC, which performs analog prefiltering of the signal before taking point-wise samples. A multiple bandpass sampling approach with $4$ channels is adopted. Each channel is composed of a crystal filter, with bandwidth of 80 KHz and extremely narrow transition bands, and is then sampled at a rate of 250 kHz. The front-end samples $4$ distinct bands of radar signal spectral content, yielding 320 Fourier coefficients after digital processing, with a total sampling rate of 1 MHz. The samples are fed into the chassis controller and a MATLAB\textsuperscript{\textregistered} function is launched that computes the 320 Fourier coefficients via FFT, composed of 4 groups of 80 consecutive Fourier coefficients. These are then used for digital recovery of the delay-Doppler map using the Doppler focusing reconstruction algorithm.

The experimental setup is based on National Instrument\textsuperscript{\textcopyright} (NI\textsuperscript{\textcopyright}) PXI equipment which is used to synthesize a radar environment and ensure system synchronization. The entire component ensemble wrapped in the NI chassis as well as the analog receiver board are depicted in Fig.~\ref{fig:proto_single}. Additional information regarding the system's configuration and synchronization can be found in \cite{baransky2014prototype}. 

To demonstrate target detection from low rate samples, the applied wave research (AWR) software is used to simulate the radar scenario, including pulse transmission and accurate power loss due to wave propagation in a realistic medium. AWR software provides a computer-based environment for the design of hardware for wireless and high speed digital products. It is used for RF, microwave and high frequency analog circuits and system design. A large variety of scenarios, consisting of different targets' parameters, i.e., delays, Doppler frequencies, and amplitudes are examined in \cite{barilan2014focusing, baransky2014prototype}. An arbitrary waveform generator (AWG) module produces an analog signal which is amplified and routed to the radar receiver board. The received radar waveform is contaminated with noise and clutter, showing the capabilities of the Xampling receiver to deal with these~\cite{barilan2014focusing, baransky2014prototype, eldar2015clutter}. The Nyquist rate of the signal is 30 MHz, so that sampling at 1 MHz corresponds to a fast time compression factor of $30$.

%%%%%%%%%%%%%%%%%%%%%%%%%%%%%%%%%%%%%%%%%%%%%%%%%%%%%%
\section{Slow Time Compression}

Most works on CS radar focus on compression in the fast time domain, reducing the number of samples per pulse below the Nyquist rate. As we have seen, using appropriate CS techniques allows preserving the range resolution while operating in low rate regimes by breaking the link between bandwidth and sampling rate. This is illustrated in Fig.~\ref{fig:res_fast} where Doppler focusing is shown to achieve the same hit rate as classic processing above a certain SNR and in Fig.~\ref{fig:screenshot1}, where close targets are seen to be correctly recovered despite sampling at $3.3\%$ of the Nyquist rate. We will now see that compression may be similarly performed in the slow time domain, as demonstrated in~\cite{cohen2016reduced}, where the number of transmitted pulses is reduced without decreasing Doppler resolution.

\subsection{Non-Uniform pulse-Doppler}
The resolution in Doppler frequency in standard processing is governed by the number of transmitted pulses $P$. More precisely, it is equal to $2 \pi /P \tau$. However, a large $P$ leads to large CPI and long time-on-target. Slow time compression breaks the relation between CPI and time-on-target. To that end, $M<P$ pulses are sent non-uniformly over the entire CPI $P \tau$, implementing non-uniform time steps between the pulses \cite{cohen2016reduced}. This way, the same CPI is kept but a smaller number of pulses is transmitted, reducing power consumption. In addition, the periods of time where no pulse is transmitted in a certain direction can be exploited to send pulses in others. This allows the radar to scan several directions at the same time and obtain the corresponding delay-Doppler maps in a single CPI. However, note that, at the same time, this reduces SNR as less pulses are transmitted in each direction.

Consider a non-uniform pulse-Doppler radar such that the $p$th pulse is sent at time $m_p \tau$, where $\{ m_p\}_{p=0}^{M-1}$ is an ordered set of integers satisfying $m_p \geq p$. In this case, (\ref{eq:uni_model}) becomes
\begin{equation}
x_T(t)= \sum_{p=0}^{M-1} h(t-m_p\tau), \quad 0 \leq t \leq P \tau,
\end{equation}
and the received frames (\ref{eq:one_frame}) are written as
\begin{equation}
\label{eq:one_frame_n}
x_p(t)= \sum_{l=0}^{L-1} \alpha_l h(t-\tau_l - m_p\tau) e^{-j \nu_l m_p \tau}, \quad 0 \leq t \leq P \tau.
\end{equation}

The same Xampling based method is used as in \cite{barilan2014focusing} to obtain the Fourier coefficients $c_p[k]$ of the received pulses. Suppose we limit ourselves to the Nyquist grid, as before, so that $\tau_l/\tau=s_l/N$, where $s_l$ is an integer satisfying $0 \leq s_l \leq N-1$, and $\nu_l \tau = 2 \pi r_l/M$, where $r_l$ is an integer in the range $0 \leq r_l \leq M-1$. Similarly to the derivations in the previous section, we can write the Fourier coefficients $c_p[k]$ in matrix form as (see \cite{cohen2016reduced}):
\begin{equation}
\label{eq:mat}
\mathbf{X}=\mathbf{HF}_N^K\mathbf{A}\left( \mathbf{F}_P^M \right)^T,
\end{equation}
where $\bf H$ is a diagonal matrix that contains the Fourier coefficients $H[k]$. The partial Fourier matrix $\mathbf{F}_M^P$ contains $M$ rows from the $P \times P$ Fourier matrix, indexed by the values of the transmitted pulses $m_p, 1 \leq p \leq M$; when sampling at the Nyquist rate, $K=N$ and $\mathbf{F}_N^K$ becomes the standard $N \times N$ Fourier matrix. Similarly, when considering uniformly spaced pulses $M=P$ and $\mathbf{F}_P^M$ is the standard $P \times P$ matrix. The goal is to recover the sparse matrix $\bf A$, that contains the values $\alpha_l$ at the $L$ indices $\{s_l,r_l\}$, from the Fourier coefficients matrix $\bf X$. 

\begin{figure}[!h]
\begin{center}
\includegraphics[width=0.5\textwidth]{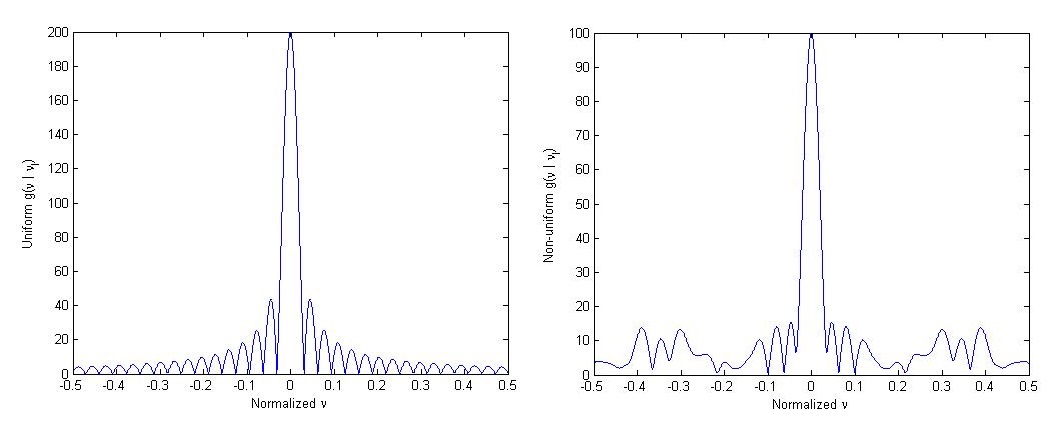}
\caption{Sum of exponents $|g(\nu|\nu_l)|$ for $M=200$, $\tau=1$sec and $\nu_l=0$ in the uniform (left) and non-uniform (right) cases. In the non-uniform case, $P=100$ pulses are chosen uniformly at random \cite{cohen2016reduced}.}
\label{fig:sum_exp}
\end{center}
\end{figure}

CS matrix recovery algorithms are directly applicable to (\ref{eq:mat}) by extending CS techniques presented in vector form, such as orthogonal matching pursuit (OMP) or the fast iterative shrinkage thresholding algorithm (FISTA) \cite{CSBook, SamplingBook, eldar2010convex} to matrix settings \cite{cs_mat_yonina}. Alternatively, instead of solving the matrix problem of (\ref{eq:mat}), we can apply the Doppler focusing operation \cite{barilan2014focusing} described in ``Doppler Focusing". As illustrated in Fig.~\ref{fig:sum_exp}, the approximation from (\ref{eq:focus_approx}) may still be applied in the non-uniform case, where $m_p \geq p$. Therefore, we can rewrite the Fourier coefficients from (\ref{eq:focused_coeff}) by replacing $p$ by $m_p$ for the non-uniform case. These may then be approximately expressed in vector form as in (\ref{eq:doppler_foc}) and recovered as previously described. It is shown in~\cite{cohen2016reduced} that the minimal number of non-uniform pulses required to recover the Doppler frequencies of $L$ targets is identical to the uniform case, that is, $2L$ pulses.

\subsection{Hardware Simulation}
\begin{figure}[!h]
\begin{center}
\includegraphics[width=0.5\textwidth]{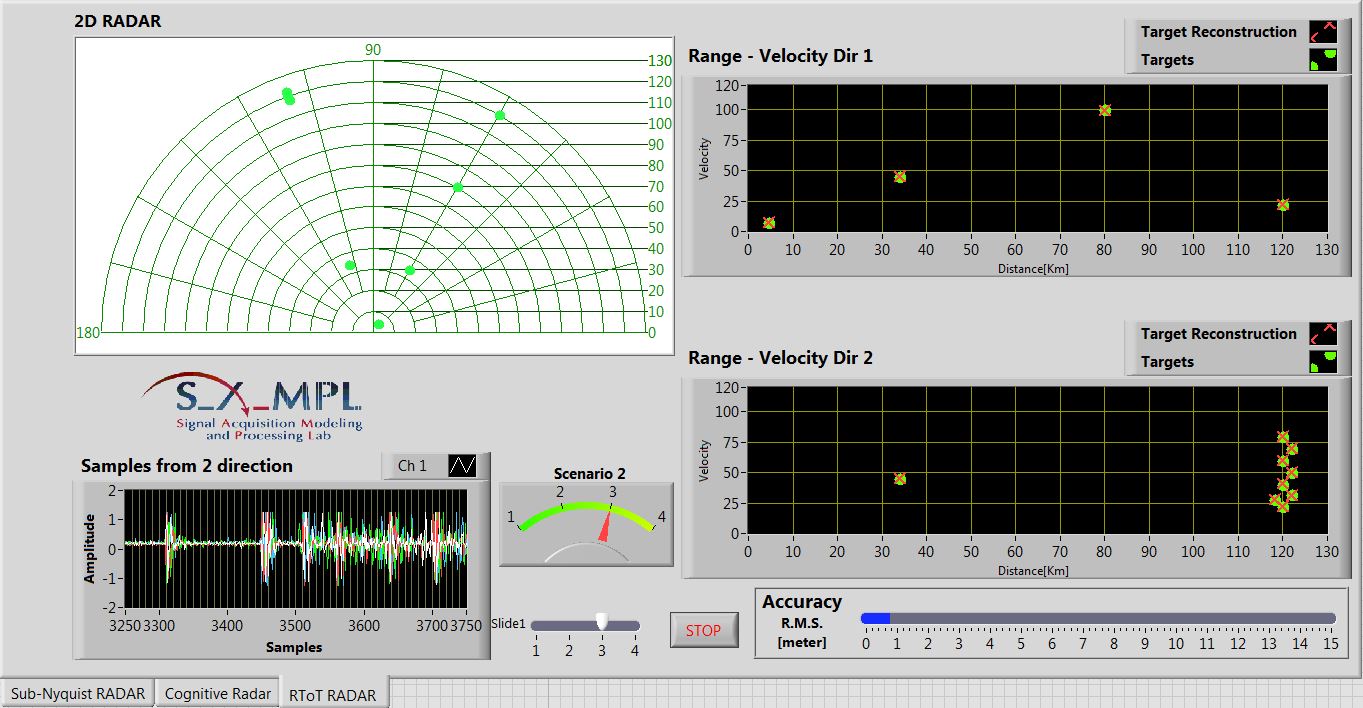}
\caption{Xampling radar, with both fast and slow time compression, experimental interface. On the left pane, true targets range in two directions (top), superposed low rate samples from both directions (bottom). On the right pane, range-velocity map of true and recovered targets in both directions \cite{cohen2016reduced}.}
\label{fig:hardware}
\end{center}
\end{figure}

Transmission of non uniform pulses has been implemented in the Xampling prototype \cite{baransky2014prototype}. Recall that the received signal has a bandwidth of $30$ MHz and is sampled at the rate of $1$ MHz. To this fast time compression, we now add compression in the slow time domain. In the hardware simulation, $P=50$ pulses over a CPI of $MP \tau=2.5 \text{sec}$ are considered. Half of the pulses, namely $M=25$, chosen at random, are sent in one direction, while the other half are sent in a second direction. Two delay-Doppler maps are then simultaneously recovered, as shown in Fig.~\ref{fig:hardware}, during a single CPI. Both maps are fully recovered, as before, from compressed samples in both the fast and slow time domains. 

\section{Range-Velocity Ambiguity Resolution}
As presented so far, targets are traditionally assumed to lie in the radar unambiguous range-velocity region. For a given PRI $\tau$, the maximum unambiguous range is $r_{\text{max}}=c \tau/2$, and the maximum unambiguous velocity is $\dot{r}_{\text{max}}=\lambda/(4\tau)$, where $\lambda$ is the radar wavelength.
When the target range and velocity intervals of interest are large, traditional pulse-Doppler radar systems suffer from the so-called ``Doppler dilemma" \cite{doviak2014doppler}, a trade-off between range and velocity ambiguity whose product is limited to $r_{\text{max}}\dot{r}_{\text{max}} =c\lambda/8$. 

Several techniques have been proposed over the years to mitigate the range-velocity ambiguity by increasing either of these parameters. Two main PRF variation based methods are staggered PRFs and multiple PRFs (MPRF). Staggered PRFs are essentially used to raise the first blind speed $\dot{r}_{\text{max}}$ significantly without degrading the unambiguous range \cite{richards2005fundamentals}. Pulse-to-pulse stagger varies the PRF from one pulse to the next, achieving increased Doppler coverage \cite{ferrari1995staggered}. The main disadvantage of this approach is that the data corresponds to a nonuniformly sampled sequence, making it more difficult to apply coherent Doppler filtering \cite{richards2005fundamentals}. In addition, clutter cancellation becomes more challenging and the sensitivity to noise increases \cite{venkatesh2016frequency}. Thus, MPRF techniques are typically preferred. We now review some of the MPRF-based methods and then present a Xampling approach that solves the delay-Doppler ambiguity using phased coded transmit pulses.

\subsection{MPRF}
The MPRF approach transmits several pulse trains, each with a different PRF. Ambiguity resolution is typically achieved by searching for coincidence between either unfolded Doppler or delay estimates for each PRF. A popular approach, adopted in \cite{ludloff1985reliability}, relies on the chinese remainder theorem \cite{skolnik} and uses two PRFs, such that the numerator and denominator of the ratio between them are prime numbers. The ambiguous velocities are computed for each train $i$, as
\begin{equation}
\hat{\dot{r}}_{i,k}=\hat{\dot{r}}_{i,0}+k \frac{\lambda}{2\tau}, \quad k \in \mathbb{Z},
\end{equation}
where $\hat{\dot{r}}_{i,0}$ is the velocity estimate within the unambiguous velocity interval $(-\dot{r}_{\text{max}}, \dot{r}_{\text{max}}]$.
Congruence between these are found by exhaustive search, so that all $\hat{\dot{r}}_{i,k}$ fall within a small interval, or correlation bin. The resulting velocity estimate is computed by averaging over all $\hat{\dot{r}}_{i,k}$.
%\begin{equation}
%\hat{\dot{r}}=\frac{1}{T} \sum_{i=1}^T \hat{\dot{r}}_{i,k_i},
%\end{equation}
%where $k_i \in \mathbb{Z}$ is the ambiguity index corresponding to the $i$th train.
Assuming $T=2$ pulse trains with PRFs with ratio $\tau_1/\tau_2=m/n$ where $m$ and $n$ are relatively prime numbers, the expanded velocity interval is of size $m\lambda/2\tau_1=n\lambda/2\tau_2$. However, in this approach, a small range error on a single PRF can cause a large error in the resolved range with no indication that this has happened \cite{trunk1993range}. 

A clustering algorithm proposed in \cite{trunk1993range} implements the search for a matching interval by computing average distances to cluster centers. The average squared error is defined as
\begin{equation}
C(k)=\sum_{i=1}^T |\hat{r}_{i,k}-\bar{r}_k|^2, \quad k=0, \dots, r_{\text{amb}}/r_{\text{max}},
\end{equation}
where $\bar{r}_k$ is the median value of the $T$ ranges with index $k$ and $r_{\text{amb}}$ is the maximal ambiguous range. The best cluster occurs at the value of $k$ where $C(k)$ is minimized. This happens when all the ambiguous ranges are unfolded correctly and hence all range estimates nearly have the same range. This technique still requires exhaustive search over clusters and does not process the samples jointly, decreasing SNR. %Other approaches \cite{ferrari1995staggered, ferrari1997doppler} have been proposed, that avoids the use of matching intervals. The folded or reduced frequency is first estimated using maxmium-likelihood criterion and it is then used to estimate the ambiguity order. However, it has been demonstrated that the ambiguity-order estimation is very sensitive to the folded frequency estimation performed initially \cite{ferrari1995staggered}.

\subsection{Phased Coded Pulses}

A random pulse phase coding (PC) approach is adopted in \cite{cohen2016staggered} to increase the range unambiguous region, while preserving that of the Doppler frequency and using a single PRF. A similar technique may be used to increase the Doppler frequency unambiguous region. Random PC has been adopted in polarimetric weather radars, which exploits the inherent random phase between pulses of the popular magnetron transmitters. In this context, PC mitigates out-of-trip echoes \cite{cao2012detection}. The approach of \cite{cohen2016staggered} introduces a random phase, which differs from pulse to pulse. Joint processing of the received signals from all trains is the key to range ambiguity resolution. 

The pulse-Doppler radar transceiver sequentially transmits one modulated pulse train, consisting of $P$ equally spaced pulses. For $0 \leq t \leq P \tau$, the transmitted signal is given by
\begin{equation}
\label{eq:uni_model_pc}
x_T(t)= \sum_{p=0}^{P-1} h(t-p\tau) e^{j c[p]},
\end{equation}
where $c[p]$ is uniformly distributed in the interval $[0, 2\pi)$ and represents the phase shift of the $p$th pulse.

As opposed to the common assumption in traditional radar, the targets' time delays $\tilde{\tau}_l$ are not assumed to lie in the unambiguous time region, namely less than the PRI $\tau$, but in the ambiguous range $\tilde{\tau}_l \in [0, Q \tau)$, where $Q < P$ is the ambiguous factor defined by the targets' maximal range. For convenience, the delay $\tilde{\tau}_l$ is decomposed into its integer part (the ambiguity order) $q_l \tau$ and fractional part (the folded or reduced delay) $\tau_l$ such that
\begin{equation} \label{eq:tau_dec}
\tilde{\tau}_l = \tau_l+q_l \tau,
\end{equation}
where $ 0 \leq q_l \leq Q-1$ is an integer and $0 \leq \tau_l < \tau$.

The received signal is then
\begin{equation}
\label{eq:uni_rec1}
x_R(t)= \sum_{p=0}^{P-1} \sum_{l=0}^{L-1} \alpha_l h(t-\tilde{\tau}_l - p\tau) e^{-j 2 \pi \nu_l (p+q_l) \tau} e^{j c[p]},
\end{equation}
for $0 \leq t < (P+Q)\tau$. The main difference with traditional pulse-Doppler radar, besides the coded phase, is that the PRI index in the Doppler shift term is $p +q_l$ rather than the pulse index $p$.

The Fourier series of the received signal (\ref{eq:uni_rec1}) can be written in matrix form, similarly to (\ref{eq:mat}), and recovered using matrix CS recovery techniques (more details may be found in \cite{cohen2016staggered}). The minimal number of samples per pulse allowing to recover $\bf X$ with high probability is found to be $K > 2L$ and the minimal number of pulses $P$ is $2L+Q+2$. This method resolves a maximum unambiguous range $r_{\text{max}}=c Q\tau/2$, while preserving the maximum unambiguous velocity $\dot{r}_{\text{max}}=\lambda/(4\tau)$, increasing their product $r_{\text{max}}\dot{r}_{\text{max}}$ by a factor of $Q$, under the above conditions on the number of samples and pulses.

This approach has three main advantages. First, it improves the delay estimation with respect to MPRF methods, since it preserves the resolution of traditional pulse-Doppler radar, namely $1/B_h$, while increasing the unambiguous delay region to $Q\tau$. Second, it increases SNR by jointly processing the samples from all pulse trains, rather than matching the estimated parameters from each pulse processed separately. Finally, it provides a systematic delay-Doppler recovery method that does not involve exhaustive search. From a practical point of view, this approach does not require the use of several pulse trains with different PRFs, simplifying hardware implementation.

\section{Cognitive Radar and Spectrum Sharing}
Recently, the concept of cognitive radar (CR) \cite{haykin_cogradar}, inspired by the echo-location system of a bat, has been presented as a natural next step for traditional radar. The cognition property requires adaptive transmission and reception capabilities, namely both the transmitter and receiver are able to dynamically adjust to the environment conditions. Many interpretations of this idea have been proposed. We focus on one aspect of cognition, the dynamic and flexible adaptation to the spectral environment, allowing spectrum sharing between communication and radar systems \cite{griffiths2015radar, fitz2014towards,bernhard2010final}. The interest in such spectrum sharing radars is largely due to electromagnetic spectrum being a scarce resource and almost all services having a need for a greater access to it.

The spectrum sharing solution proposed in~\cite{cohen2016specx} capitalizes on the cognitive abilities of the radar system. It is shown how compressed radars may be adapted to allow for spectral coexistence between communication and radar signals and flexibility of the radar transmission. This demonstrates that, beyond increasing resolution and realizing compression in the time, frequency and spatial domains, compressed radars have the potential to enable otherwise challenging technologies.

\subsection{Spectral Adaptive Transmission}

In previous works that implement fast time compression, e.g., Xampling radar \cite{barilan2014focusing, baransky2014prototype}, the transmitter broadcasts a wideband signal, which reflects off the targets and propagates back to the receiver. The received signal is then filtered before sampling, so that only the content of a few narrow bands is sampled and processed. These works only deal with the reception side of the radar, providing sampling and processing techniques that can be used with any traditional radar transmitter. However, for broadband frequency occupation and power savings, only the narrow frequency bands that are to be sampled may be transmitted \cite{cohen2016towards, cohen2016specx}. This will not affect any aspect of the processing since the received signal is preserved in the bands of interest. In fact, since all the signal power is concentrated in the processed bands, the SNR increases and the detection performance improves \cite{mishra2017xampling}.

Let $\tilde{H}(f)$ be the CTFT of the new transmitted radar pulse,
\begin{equation}
\tilde{H}(f)= \left\{ \begin{array}{ll}
H(f) & f \in [f_r^i -B_r^i/2, f_r^i +B_r^i/2] \text{ for } 1 \leq i \leq N_b \\
0 & \text{else},
\end{array} \right.
\end{equation}
where $N_b$ is the number of filtered bands, $B_r^i$ and $f_r^i$ are the bandwidth and center frequency of the $i$th band, respectively. Obviously, the computation of the relevant Fourier coefficients $c_p[k]$ from (\ref{eq:fourier_coeff}) will not change. Therefore, the recovery methods presented in ``Fast Time Xampling" are applicable here as well.

The concept of transmitting only a few subbands that the receiver processes is one way to formulate a frequency agile CR, in terms of adaptation to spectral demands. Complying with CR requirements, the support of the subbands varies with time to allow for dynamic and flexible adaptation. Such a system also enables the radar to disguise the transmitted signal as an electronic counter measure (ECM) or cope with crowded spectrum by using a smaller interference-free portion, as further discussed below.

\subsection{Application to Spectrum Sharing}

The unhindered operation of a radar that shares its spectrum with communication systems has captured a great deal of attention within the operational radar community in recent years \cite{griffiths2015radar, fitz2014towards,bernhard2010final, cohen2016specx}. Recent research programs in spectrum sharing radars include the Enhancing Access to the Radio Spectrum (EARS) project by the National Science Foundation (NSF) \cite{bernhard2010final} and the Shared Spectrum Access for Radar and Communication (SSPARC) program \cite{fitz2014towards, jacyna2016ssparc} by the Defense Advanced Research Projects Agency (DARPA). 

A variety of system architectures have been proposed for spectrum sharing radars. Most put emphasis on optimizing the performance of either radar \cite{patton2012disjoint,stinco2016spectrum} or communication \cite{nartasilpa2016commrad} while ignoring the performance of the other. In nearly all cases, real-time exchange of information between radar and communication hardware has not yet been integrated into the system architectures. Exceptions to this are automotive solutions where the same waveform is used for both target detection and communication \cite{kumari2015investigating}.

In a similar vein, the sub-Nyquist CR based approach from \cite{cohen2016specx} incorporates handshaking of spectral information between the two systems. The CR configuration is key to spectrum sharing since the radar transceiver can adapt its transmission to available bands, achieving coexistence with communication signals. Suppose the set of all frequencies of the available common system spectrum is given by $\mathcal{F}$. The communication and radar systems occupy the subsets $\mathcal{F}_C$ and $\mathcal{F}_R$ of $\mathcal{F}$, respectively. The goal is to design the radar waveform and its support $\mathcal{F}_R$, conditional on the fact that the communication occupies frequencies $\mathcal{F}_C$, unknown to the radar transceiver~\cite{cohen2016specx}. To detect the bands left vacant by the communication signals, spectrum sensing needs to be performed over a large bandwidth. Such a task has recently received tremendous interest in the communication community, which faces a bottleneck in terms of spectrum availability. To increase the efficiency of spectrum managing, dynamic opportunistic exploitation of temporarily vacant spectral bands by secondary users has been considered, under the name of Cognitive Radio (CRo) \cite{mitola1999cognitive, cohen2017magazine}.

\begin{figure*}[!t]
	\centering
		\includegraphics[width=\textwidth]{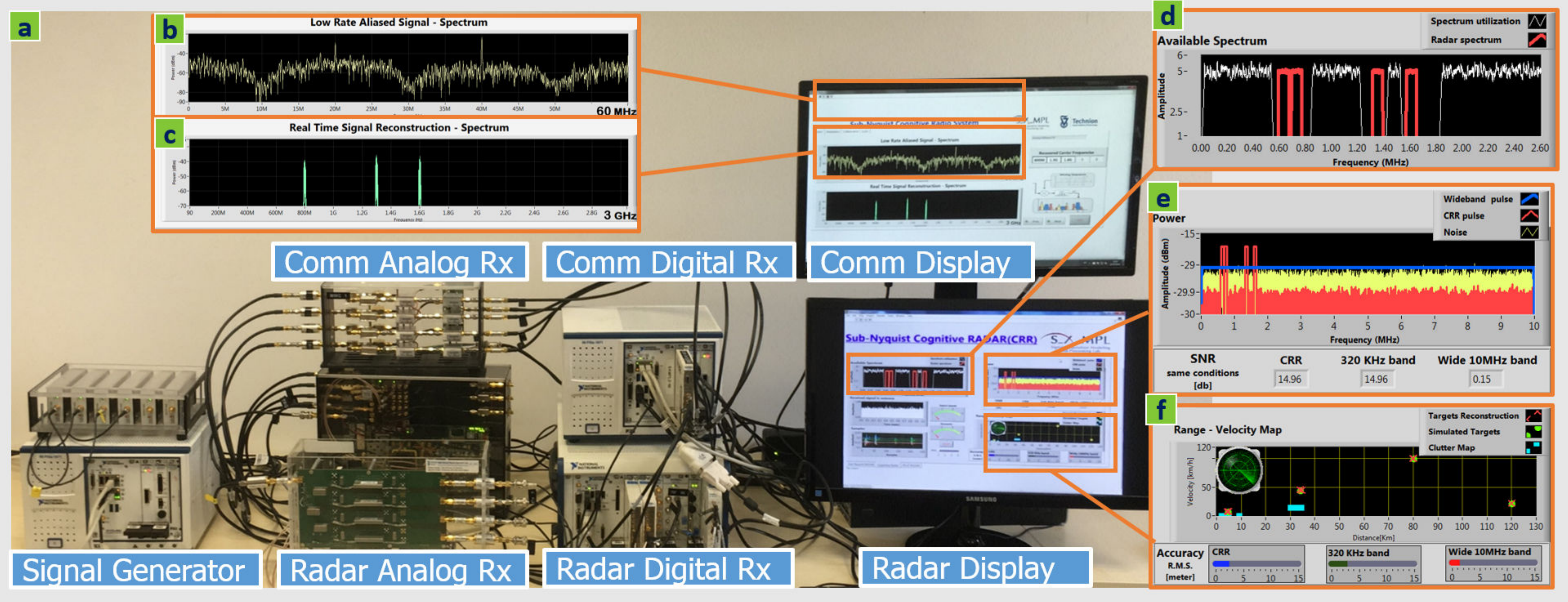}
		\caption{(a) SpeCX prototype. The system consists of a signal generator, a CRo communication analog receiver including the MWC analog front-end board, a communication digital receiver, a CR analog and receiver. SpeCX comm system display showing (b) low rate samples acquired from one MWC channel at rate $120\:\text{MHz}$, and (c) digital reconstruction of the entire spectrum from sub-Nyquist samples. SpeCX radar display showing (d) coexisting communication and CR, (e) CR spectrum compared with the full-band radar, and (f) range-velocity display of detected and true locations of the targets \cite{cohen2016specx}.} 
        \vspace{-.3cm}
		\label{fig:proto}
\end{figure*}

A spectrum sharing paradigm using Xampling techniques, the spectral coexistence via Xampling (SpeCX) system \cite{cohen2016specx} is composed of a sub-Nyquist cognitive radio (CRo) receiver (see \cite{cohen2017magazine} for more details) to detect the occupied communication bands, so that the radar transmitter may subsequently exploit the spectral holes. In this setting, the received signal at the communication receiver is given by
\begin{equation}
x(t)=x_C(t)+x_R(t),
\end{equation}
where $x_R(t)=r_{T_X}(t)+r_{R_X}(t)$ is the radar signal sensed by the communication receiver, composed of the transmitted and received radar signals. The goal is thus to recover the support of $x_C(t)$, given the known support of $x_R(t)$ which is shared by the radar transmitter with the communication receiver. This can be formulated as a sparse recovery with partial support knowledge, studied under the framework of modified CS \cite{vaswani2010modified}.

Once $\mathcal{F}_C$ is identified, the communication receiver provides a spectral map of occupied bands to the radar. Equipped with the detected spectral map and known radio environment map (REM), the objective of the radar is to identify an appropriate transmit frequency set $\mathcal{F}_R \subset \mathcal{F}\setminus\mathcal{F}_C$ such that the radar's probability of detection $P_d$ is maximized.  For a fixed probability of false alarm $P_{\text{fa}}$, the $P_d$ increases with higher signal to interference and noise ratio (SINR) \cite{kay1998fundamentals}. Hence, the frequency selection process can, alternatively, choose to maximize the SINR or minimize the spectral power in the undesired parts of the spectrum. In order to find available bands with least interference, a structured sparsity framework \cite{huang2011learning} is adopted in \cite{cohen2016specx}. Additional requirements of transmit power constraints, range sidelobe levels, and minimum separation between the bands can also be imposed. At the receiver of this spectrum sharing radar, the sub-Nyquist processing method of \cite{barilan2014focusing} recovers the delay-Doppler map from the subset of Fourier coefficients defined by $\mathcal{F}_R$.

This CR system leads to three main advantages. First, the CS reconstruction, performed as presented in \cite{barilan2014focusing} on the transmitted fragmented bands, achieves the same resolution as traditional Nyquist processing over a significantly smaller bandwidth. Second, by concentrating all the available power in the transmitted narrow bands rather than over a wide bandwidth, the CR increases SNR. Finally, this technique allows for a dynamic form of the transmitted signal spectrum, where only a small portion of the whole bandwidth is used at each transmission, enabling spectrum sharing with communication signals, as illustrated in Fig.~\ref{fig:proto}(d). There, coexistence between radar transmitted bands in red and existing communication bands in white is shown.

\subsection{SpeCX Prototype}

The SpeCX prototype, presented in Fig.~\ref{fig:proto}, demonstrates radar and communication spectrum sharing. It is composed of a CRo receiver and a CR transceiver. At the heart of the CRo system lies the proprietary modulated wideband converter (MWC) board \cite{mishali2011xampling} that implements a sub-Nyquist analog front-end receiver, which processes signals with Nyquist rates up to $6\, \text{GHz}$. The card first splits the wideband signal into $M=4$ hardware channels, with an expansion factor of $q=5$, yielding $Mq=20$ virtual channels after digital expansion (see \cite{mishali2010theory} for more details on the expansion). In each channel, the signal is mixed with a periodic sequence $p_i(t)$, which are truncated versions of Gold Codes~\cite{gold1967optimal}, generated on a dedicated FPGA, with periodic frequency $f_p=20\, \text{MHz}$.

Next, the modulated signal passes through an analog anti-aliasing LPF. Finally, the low rate analog signal is sampled by a NI\textsuperscript{\textcopyright} ADC operating at $f_s=(q+1)f_p=120\,\text{MHz}$ (with intended oversampling), leading to a total sampling rate of $480\,\text{MHz}$. The digital receiver is implemented on a NI\textsuperscript{\textcopyright} PXIe-1065 computer with DC coupled ADC. Since the digital processing is performed at the low rate of $120 \, \text{MHz}$, very low computational load is required in order to achieve real time recovery. 
The prototype is fed with RF signals composed of up to $N_{\text{sig}}=5$ real communication transmissions, namely $10$ spectral bands with total bandwidth occupancy of up to $200\,\text{MHz}$ and varying support, with Nyquist rate of $6\,\text{GHz}$.

The input transmissions then go through an RF combiner, resulting in a dynamic multiband input signal, that enables fast carrier switching for each of the bands. This input is specially designed to allow testing the system's ability to rapidly sense the input spectrum and adapt to changes, as required by modern CRo and shared spectrum standards, e.g. in the SSPARC program. The system's effective sampling rate, equal to $480\,\text{MHz}$, is only $8\%$ of the Nyquist rate. Support recovery is digitally performed on the low rate samples. The prototype successfully recovers the support of the communication transmitted bands, as demonstrated in Fig.~\ref{fig:proto}(b)-(c). Once the support is recovered, the signal itself can be reconstructed from the sub-Nyquist samples. This step is performed in real-time, reconstructing the signal bands one sample at a time.

The CR receiver system is identical to the sub-Nyquist sampling prototype of \cite{barilan2014focusing, cohen2016towards, baransky2014prototype}. In the cognitive case, the transmitter only transmits over $N_b=4$ bands, which constitute $3.2\%$ of the original wideband signal bandwidth, after the spectrum sensing process has been completed by the communication receiver. Figure~\ref{fig:proto}(d) illustrates coexistence between the radar transmitted bands in red and the existing communication bands in white. The gain in power is demonstrated in Fig.~\ref{fig:proto}(e); the wideband radar spectrum is shown in blue, the CR in red and the noise in yellow on a logarithmic scale. The true and recovered range-velocity maps are presented in Fig.~\ref{fig:proto}(f). All $L=10$ targets are perfectly recovered and clutter, depicted in blue, is discarded. Below the map, the range recovery accuracy is shown for 3 scenarios: from left to right, CR in blue ($2.5$m), 4 adjacent bands with same bandwidth ($12.5$m) and wideband ($4$m). The second configuration selects 4 adjacent frequency bands with the same bandwidth as the CR (with non adjacent bands) for transmission. Its poor resolution stems from its small aperture. The CR system with non-adjacent bands yields better resolution than traditional wideband transmission, sampling and processing at the Nyquist rate, due to the increased SNR.

\section{Compressed MIMO Radar}

Compressed radar methods have recently been extended to MIMO settings, where their impact may be even greater than for single antenna configurations. MIMO radar systems belong to the family of array radars, which allow recovering simultaneously the targets' ranges, Dopplers and azimuths. This three-dimensional recovery results in high digital processing complexity. One of the main challenges of MIMO radar is therefore coping with complicated systems in terms of cost, high computational load and hardware implementation. CS has thus naturally been applied to MIMO in order to reduce the processing complexity on the digital side, as well as allow for spatial compression, in addition to time compression achieved in single antenna systems. In MIMO radars, the array aperture, which depends on the number of antennas, dictates the azimuth resolution. Since the aperture is determined by the number of antennas in traditional virtual ULAs, high azimuth resolution requires a large number of antennas.

\subsection{Increased Resolution}

As in single antenna radar systems, CS has first been exploited to increase parameter resolution. Here, the MIMO array is composed of $T$ transmitters and $R$ receivers so as to achieve the desired aperture $Z=\frac{TR}{2}$, as shown in Fig.~\ref{fig:arrays1}. The transmitted signal at the $m$th transmit antenna is given by (\ref{trMth}) and each receiver samples the received signal at the Nyquist rate, as in traditional MIMO. Assuming a sparse target scene, where the ranges, Dopplers and azimuths lie on a predefined grid, the work of \cite{herman2009high} is extended to MIMO architectures in \cite{strohmersparse, strohmersparse2}. The transmit and receive array manifolds are respectively given by
\begin{equation}
\mathbf{a}_T(\theta)=[e^{j2 \pi \xi_1 \theta}, e^{j2 \pi \xi_2 \theta}, \dots, e^{j2 \pi \xi_{T} \theta}]^T,
\end{equation}
and 
\begin{equation}
\mathbf{a}_R(\theta)=[e^{j2 \pi \zeta_1 \theta}, e^{j2 \pi \zeta_2 \theta}, \dots, e^{j2 \pi \zeta_R \theta}]^T,
\end{equation}
where $\xi_m$ and $\zeta_q$ are the relative $m$th transmit and $q$th receive antenna spacings. The $R \times N$ received signal matrix from a unit strength target at direction $\theta$, with delay $\tau$ and Doppler $\nu$ is defined as
\begin{equation}
\mathbf{Z}=\mathbf{a}_R(\theta)\mathbf{a}_T^T(\theta) \mathbf{S}^T(\tau,\nu).
\end{equation}
Here, $\mathbf{S}^T(\tau,\nu)_{i,m}=s_m(t_i-\tau)e^{j 2 \pi \nu t_i}$ where $t_i$ are the sampling times and $m$ indexes the transmitted waveforms. In this case, the columns of the dictionary $\bf A$ are given by $\text{vec}(\mathbf{Z})$ for all possible combinations of $\theta$, $\nu$ and $\tau$ on a predefined grid.

The targets' parameters are recovered by matching the received signal with dictionary atoms. To achieve measurement diversity, random waveforms may be used, while the antenna locations are deterministic. ULAs are considered in \cite{strohmersparse}, for both the transmit and receive arrays, that do not benefit from the virtual array configuration. Alternatively, deterministic waveforms can be used, such as Kerdock codes \cite{strohmersparse2} while the antenna locations are selected uniformly at random over the aperture $Z=\frac{TR}{2}$. Bounds on $N$ with respect to the number of antennas $T$ and $R$ and the number of samples, that ensure targets' parameters recovery, are provided in \cite{strohmersparse, strohmersparse2}.

A similar approach extends the framework of \cite{herman2009high} to the MIMO setting by adding an azimuth matrix to the time-shift and frequency modulation matrices $\bf T$ and $\bf M$, respectively, defined in (\ref{eq:matTandM}). In this case, each target lying on the grid is represented by a time-shift, a frequency modulation and an angle $\mathbf{A}_{q,m}=e^{j\theta(\xi_m+\zeta_q)}$ \cite{chen2008compressed}.

In both works, assuming $N$ grid points in each dimension, the number of columns of $\bf A$ is $N^3$. The processing efficiency is thus penalized by a very large dictionary that contains every parameter combination. Note that the above works focus on increased parameter resolution and do not deal with reduced time/spatial sampling and processing rates.

\subsection{Reduced Processing}

Fast time compression is performed in \cite{yu2010mimo, kalogerias2014matrix, mimoMC}, where the Nyquist rate samples are compressed in each antenna before being forwarded to the central unit. A circular array is adopted in \cite{yu2010mimo}, with transmit and receive nodes uniformly distributed on a disk with small radius. At each receive antenna, linear projections of the measurement vector are retained so that the resulting samples are compressed in both the slow and fast time domains. Both individual reconstruction at each receiver and joint processing at a fusion center are proposed, using CS recovery methods. The actual sampling is still performed at the Nyquist rate.

The MIMO matrix completion (MIMO-MC) radar \cite{kalogerias2014matrix, mimoMC} employs matrix completion techniques in order to avoid parameter discretization, typically used in CS methods. Two configurations are proposed for azimuth-Doppler recovery in a range bin of interest. In the first scenario, each receiver performs MF and forwards the maximum of each MF output to the fusion center. The samples from the $p$th pulse transmitted to the fusion center can then be written in matrix form as
\begin{equation}
\mathbf{X}_p=\mathbf{A}_R \mathbf{\Sigma} \mathbf{D}_p \mathbf{A}^T_T,
\end{equation}
where $\mathbf{X}_p$ is the $R \times T$ matrix that contains the maximum of the MF output for each transmitter and each receiver. For ULA configurations, the $l$th column of the $T \times L$ transmitter steering matrix $\mathbf{A}_T$ is given by $(\mathbf{A}_T)_l = [1, e^{j\frac{2\pi}{\lambda} d_T \sin(\theta_l)}, \dots , e^{j\frac{2\pi}{\lambda} (T-1)d_T \sin(\theta_l)}]^T$, where $d_T$ is the inter-element spacing. The steering matrix $\mathbf{A}_R$ at the receiver is similarly defined. The diagonal matrix $\bf \Sigma$ contains the targets' RCS $\alpha_l$ and the diagonal matrix $\mathbf{D}$ contains the targets' Dopplers such that ${\mathbf{D}_p}_{(l,l)}=e^{j\frac{2\pi}{\lambda} 2 \nu_l (p-1)\tau}$. In this scheme, each receiver transmits the output of a few randomly chosen MFs to the fusion center so that $\mathbf{X}_p$ is only partially known. 

In the second scenario, the receivers forward Nyquist samples to the fusion center, without performing MF. In this case, the samples are written as
\begin{equation}
\mathbf{X}_p=\mathbf{A}_R \mathbf{\Sigma} \mathbf{D}_p \mathbf{A}^T_R \mathbf{S},
\end{equation}
where $\bf S$ is the $T \times N$ matrix that contains the Nyquist rate samples of each transmitted waveform $s_m(t)$.
In this scheme, each receive antenna randomly acquires a subset of the Nyquist samples, and transmits these to the fusion center. In both cases, the fusion center performs MC before parametric estimation methods are applied to extract $\theta_l$ and $\nu_l$, such as MUSIC \cite{pisarenko73, schmidt86}. In these works, sampling and processing rate reduction are not addressed since compression is performed in the digital domain, after sampling, and the missing samples are reconstructed before recovering the targets' parameters. Instead, these approaches are aimed at reducing the communication overhead between the receivers and the fusion center.

\subsection{Spatial Compression}
\begin{figure}[!h]
\begin{center}
\includegraphics[width=0.5\textwidth]{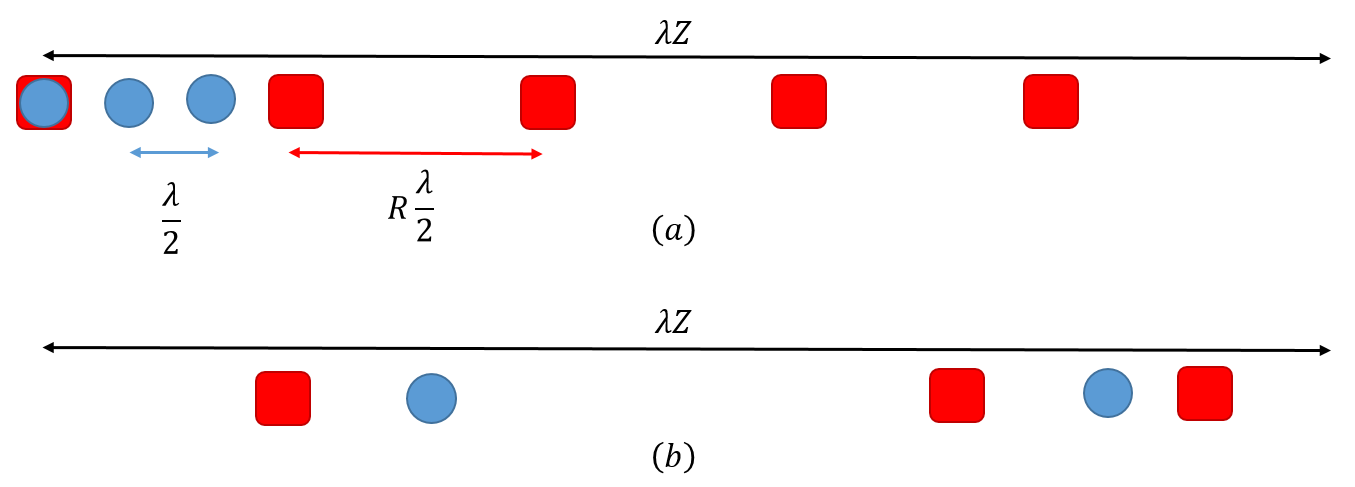}
\caption{Illustration of MIMO arrays: (a) standard array, (b) random thinned array \cite{cohen2016summer}.}
\label{fig:arrays2}
\end{center}
\end{figure}

Several recent works have considered applying CS to MIMO radar to reduce the number of antennas or the number of samples per receiver without degrading resolution. The problem of azimuth recovery of targets all in the same range-Doppler bin is investigated in \cite{rossi2014spatial}. Spatial compressive sampling is performed, where the number of antennas is reduced while preserving azimuth resolution. The classic MIMO virtual array configuration requires receivers with maximum spacing $\lambda/2$ and transmitters with spacing $R\lambda/2$ (or vice versa). The product $RT$ thus scales linearly with aperture which sets the azimuth resolution. Spatial compression is achieved by using a sparse random array architecture \cite{rossi2014spatial}, in which a low number of transmit and receive elements are placed at random over the same aperture $Z$, achieving similar resolution as a filled array, but with significantly fewer elements. The random array configuration is illustrated in Fig.~\ref{fig:arrays2}. Beamforming is applied on the time domain samples obtained from the thinned array at the Nyquist rate and the azimuths are recovered using CS techniques. Recovery guarantees and guidelines concerning the choice of the product $RT$ and the antenna locations are provided. Methods for choosing the antenna locations using deep networks are investigated in~\cite{elbir2018cognitive}.

\subsection{Time and Spatial Compression}
In all the above works, recovery is performed in the time domain on acquired or reconstructed Nyquist rate samples for each antenna. The sub-Nyquist MIMO radar (SUMMeR) system, presented in~\cite{cohen2016summer}, extends the Xampling concept to MIMO configurations and breaks the link between aperture and number of antennas, similarly to \cite{rossi2014spatial}. The concept of Xampling is applied both in space (antenna deployment) and in time (sampling scheme) in order to simultaneously reduce the required number of antennas and samples per receiver, while preserving time and spatial resolution. In particular, targets' azimuths, ranges and Dopplers are recovered from compressed samples in both space and time, while keeping the same resolution induced by Nyquist rate samples obtained from a full virtual array with low computational cost.

\begin{figure}[!h]
\begin{center}
\includegraphics[width=0.45\textwidth]{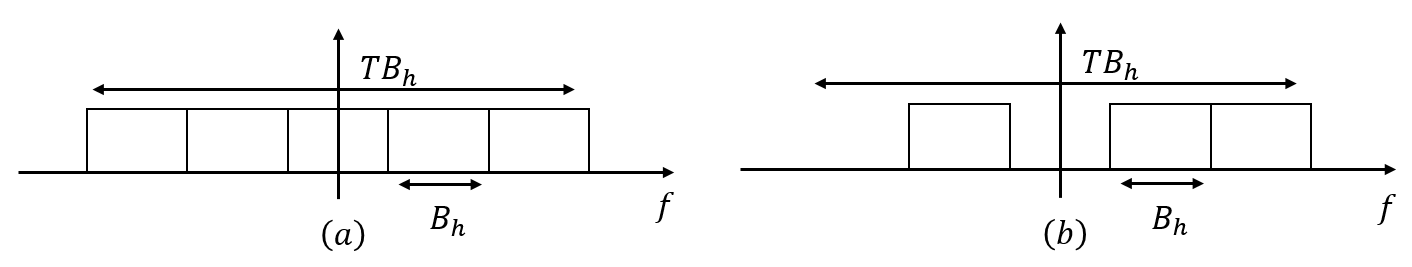}
\caption{FDMA transmissions: (a) standard, (b) spatial compression \cite{cohen2016summer}.}
\label{fig:specs}
\end{center}
\end{figure}

%% Thinned array
The SUMMeR system implements a collocated MIMO radar system with $M<T$ transmit antennas and $Q<R$ receive antennas, whose locations are chosen uniformly at random within the aperture of the virtual array described above, that is 
$\{\xi_{m}\}_{m=0}^{M-1} \sim \mathcal{U} \left[ 0, Z \right]$ and $\{\zeta _{q}\}_{q=0}^{Q-1} \sim \mathcal{U} \left[ 0,Z \right]$, respectively, as shown in Fig.~\ref{fig:arrays2}. Note that, in principle, the antenna locations may be chosen on the ULAs' grid. However, this configuration is less robust to range-azimuth ambiguity and leads to coupling between these parameters in the presence of noise \cite{cohen2016summer}. An FDMA framework is adopted so that spatial compression, which in particular reduces the number of transmit antennas, removes the corresponding transmitting frequency bands as well. The transmitted signals are illustrated in Fig.~\ref{fig:specs} in the frequency domain. Figure~\ref{fig:specs}(a) and (b) show a standard FDMA transmission for $T=5$ and the resulting signal after spatial compression for $M=3$.

The transmitted pulses, defined in (\ref{trMth}), are reflected by the targets and collected at the receive antennas. Under the assumptions described in ``Targets' Assumptions", the received signal $\tilde{x}_q(t)$ at the $q$th antenna is a sum of time-delayed, scaled replica of the transmitted signals:
    \begin{equation}
    {\tilde{x}_{q}}\left( t \right) =  \sum\limits_{m=0}^{T-1} \sum\limits_{l=1}^{L} {\alpha_l {{s}_{m}}\left( \frac{c+v_l}{c-v_l} \left(t-\frac{R_{l,mq}}{c+v_l} \right)  \right)},
    \end{equation}
where $R_{l,mq}$ is the sum of the distances from the $m$th transmitter and $q$th receiver to the $l$th target, accounting for the array geometry. After demodulation to baseband, the received signal can be further simplified to
\begin{equation} \label{eq:rec_sig}
 x_q \left( t \right) = \sum\limits_{p=0}^{P-1} \sum\limits_{m=0}^{M-1} \sum\limits_{l=1}^{L}  \alpha_l h_m \left( t- p\tau  - \tau _{l} \right) e^{j2 \pi \beta_{mq} \vartheta _l} e^{j 2\pi f^D_l p \tau},
\end{equation}
where $\beta_{mq} = \left( \zeta _{q}+\xi _m \right) \left( f_m \frac{\lambda}{c} +1 \right)$, with $f_m$ the $m$th transmission carrier frequency and $\lambda$ the signal wavelength.
%It is convenient to express $x_q(t)$ as a sum of single frames
%\begin{equation}
%\label{eq:frames_mimo}
%x_q(t)= \sum_{p=0}^{P-1} x_q^p(t),
%\end{equation}
%where
%\begin{equation}
%\label{eq:one_frame_mimo}
%x_q^p(t)= \sum_{m=0}^{M-1} \sum_{l=1}^{L} \alpha_l h(t-\tau_l - p\tau) e^{j2 \pi \beta_{mq} \vartheta _l} e^{j 2\pi f^D_l p \tau}.
%\end{equation}
The goal is to estimate the targets' ranges, azimuths and velocities, i.e. to estimate ${{\tau}_{l}}$, ${{\vartheta}_{l}}$ and $f^D_l$ from low rate samples of $x_q(t)$, and small numbers $M$ and $Q$ of antennas.

Similarly to the Xampling processing in \cite{barilan2014focusing}, SUMMeR considers the Fourier coefficients of the received signal $x_q^p(t)$ at the $q$th antenna. To jointly recover the targets' ranges, azimuths and Doppler frequencies, the concept of Doppler focusing from \cite{barilan2014focusing} (see ``Doppler Focusing") is applied to the MIMO setting and CS algorithms are extended to simultaneous matrix recovery \cite{cohen2016summer}. The minimal number of channels required for perfect recovery of $\mathbf{X}_D$ with $L$ targets in noiseless settings is $MQ \geq 2L$ with a minimal number of $MK \geq 2L$ samples per receiver and $P \geq 2L$ pulses per transmitter \cite{cohen2016summer}. The SUMMeR system has been implemented in hardware, as described in the next section.

\subsection{Hardware Prototype}

\begin{figure*}[t]
\centering
\includegraphics[width=\textwidth]{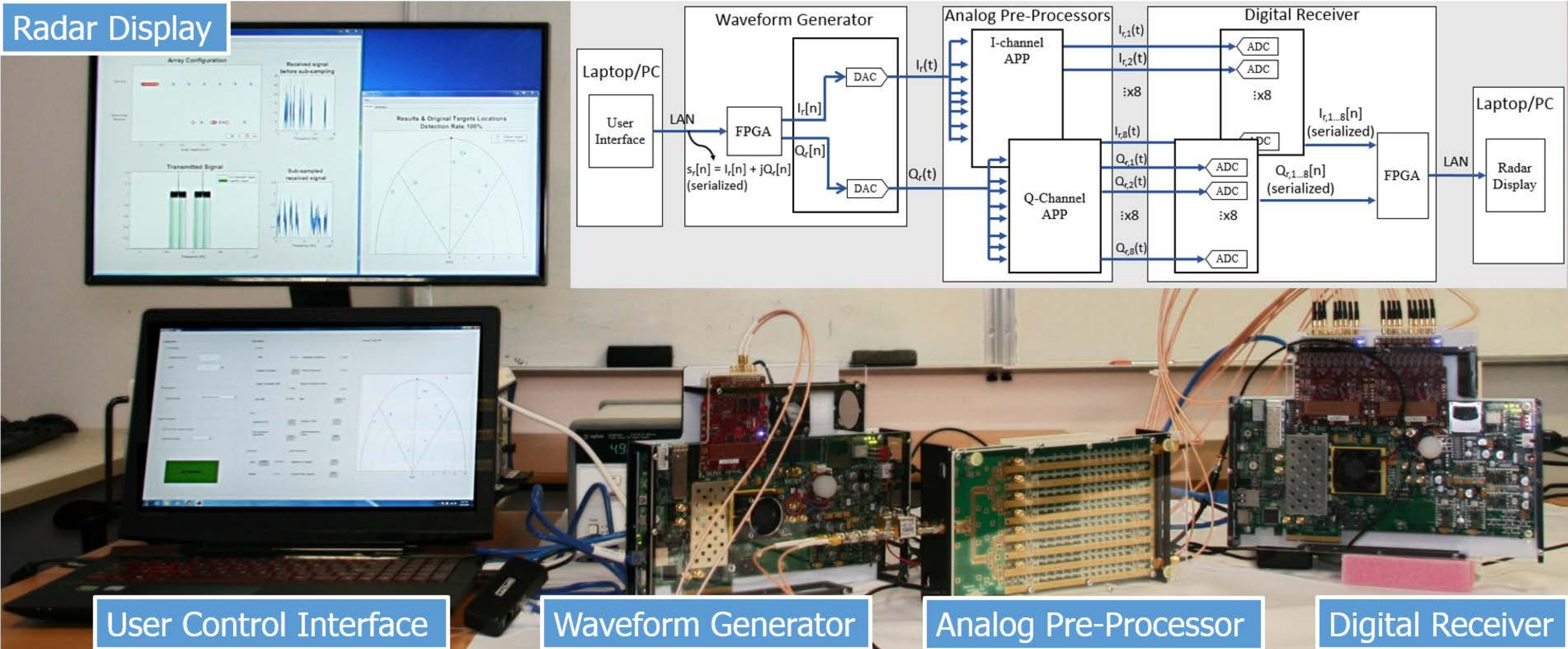}
\caption{Sub-Nyquist MIMO prototype and user interface. The analog pre-processor (APP) module consists of two cards mounted on opposite sides of a common chassis. The inset shows the simplified block diagram of the system. The subscript $r$ represents received signal samples for $r$th receiver. Wherever applicable, the second subscript corresponds to a particular transmitter. The square brackets (parentheses) are used for digital (analog) signals \cite{mishra2016cognitive, cohen2016radarconmimo}.}
\label{fig:mimoblockphoto}
\end{figure*}

The cognitive SUMMeR prototype \cite{mishra2016cognitive, cohen2016radarconmimo} extends the Doppler focusing Xampling based prototype~\cite{baransky2014prototype} to the MIMO configuration. It simultaneously recovers the targets' delays, Dopplers and azimuths from sub-Nyquist samples. More specifically, it implements a receiver with a maximum of 8 transmit (Tx) and 10 receive (Rx) antenna elements. The same hardware is used for each receive element and serially feeds the signals of all $R=10$ receivers to the same prototype.

In order to avoid use of an overwhelmingly large number of ADCs and bandpass filters for an $8 \times 10$ array, a cognitive transmission is adopted wherein each transmit signal lies in $N_b=8$ disjoint, narrow slices, over a $15$ MHz band. Each subband is of width $375$ kHz, leading to a total signal bandwidth of $3$ MHz. The transmit subbands locations were chosen so that all can be sub-sampled using a single low-rate ADC without aliasing between them~\cite{cohen2014channel}. This allows reducing the number of samplers. The signal is subsampled at 7.5 MHz whereas a non-cognitive signal would have occupied the entire $15$ MHz spectrum requiring a Nyquist sampling rate of $30$ MHz. Therefore, the use of cognitive transmission enables spectral sampling reduction by a factor of $4$ for each channel. The effective signal bandwidth is reduced by a factor of $5$ ($=15$ MHz$/3$ MHz) respectively for each channel.

The system may be configured to operate in various array configurations simulating different numbers and locations of the antennas. The hardware switches off the inactive channels and does not sample any data over the corresponding ADCs. This governs the spatial compression by reducing the amount of receivers and transmitters. In its baseline configuration, the system uses only half the antennas with respect to the full virtual array, that is $M=4$ transmitters and $Q=5$ receivers. Figure~\ref{fig:mimoblockphoto} shows the sub-Nyquist MIMO prototype, user interface and radar display. The inset graph depicts the signal flow through a simplified block diagram.

The experimental process consists of the following steps. The simulated radar scenario is stored in a custom-designed waveform generator. The scenario includes modeling of pulse transmission, accurate power loss due to wave propagation in a realistic medium, and interaction of a transmit signal with the target. A large variety of scenarios, consisting of different targets' parameters, i.e., delays, Doppler frequencies, and amplitudes, and array configurations, i.e. number of transmitters and receivers and antenna locations, may be examined using the prototype. The waveform generator board then produces an analog signal corresponding to the synthesized radar environment, which is amplified and routed to the MIMO radar receiver board. The prototype samples and processes the signal in real-time. The physical array aperture and simulated target response correspond to an X-band ($f_c = 10$ GHz) radar.

\begin{figure}[!h]
\begin{center}
\includegraphics[width=0.5\textwidth]{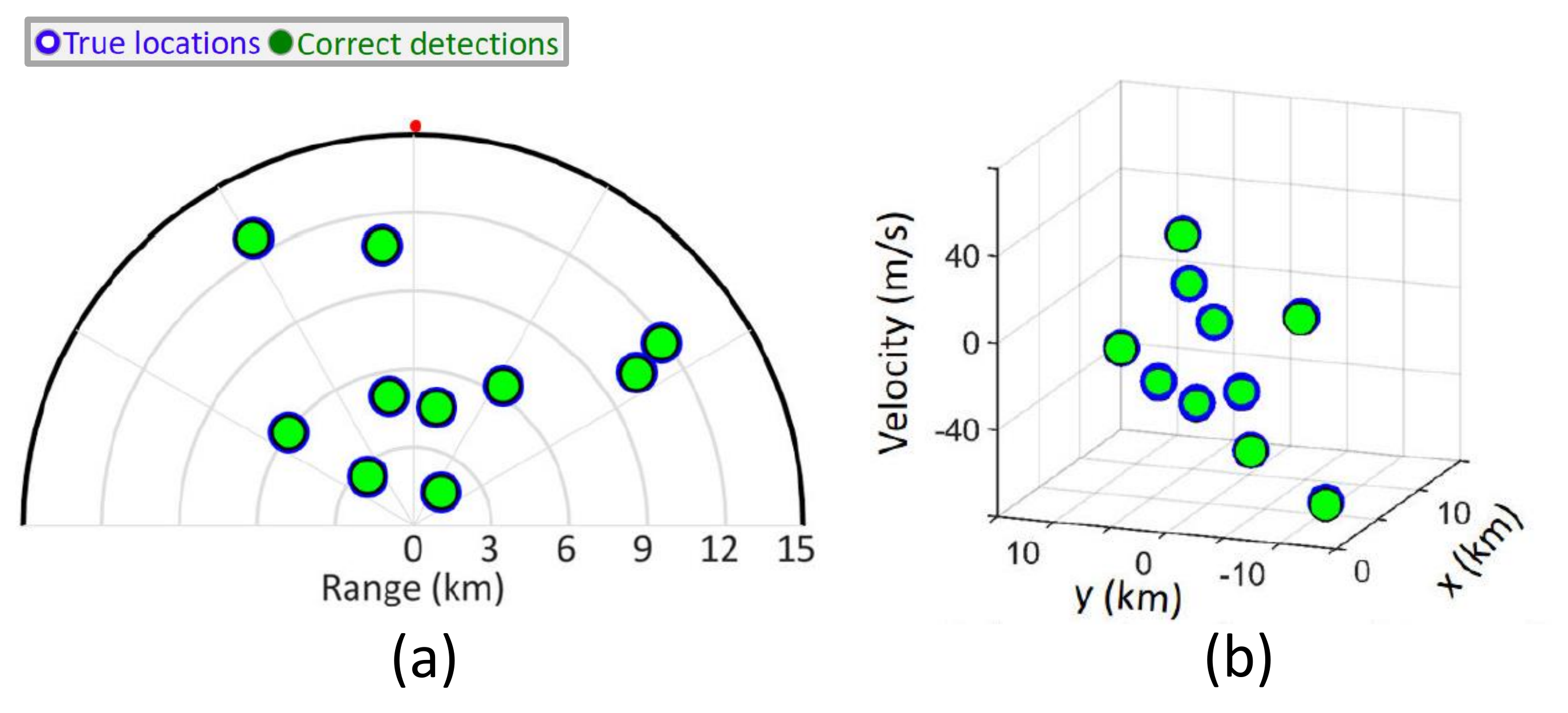}
\caption{SUMMeR prototype recovery performance: (a) Plan Position Indicator (PPI) display. The origin is the location of the radar. The red dot indicates the north direction relative to the radar. Positive (negative) distances along the horizontal axis correspond to the east (west) of the radar. Similarly, positive (negative) distances along the vertical axis correspond to the north (south) of the radar. The estimated targets are plotted over the ground truth. (b) Range-Azimuth-Doppler map for the same targets. The lower axes represent the cartesian coordinates of the polar representation of the PPI plots from (a). The vertical axis represents the Doppler spectrum~\cite{cohen2016summer}.}
\label{fig:mimo_sim}
\end{center}
\end{figure}

Figure~\ref{fig:mimo_sim} presents some recovery results from the prototype. In the experiment, $P=10$ pulses were transmitted at a uniform PRF of $100\mu$Hz. The received signal corresponding to the echoes from $L=10$ targets, placed at arbitrary range and azimuths and with arbitrary velocities, was injected into the transmit waveform generator.  In the experiment, when the angular spacing (in terms of the sine of azimuth) between any two targets was greater than $0.025$ and the signal SNR = $-8$ dB, the recovery performance of the compressed configuration in time and space was equivalent to that of a full array, that is with $8$ transmitters and $10$ receivers. The figure shows the obtained plan position indicator (PPI) plot and range-azimuth-Doppler maps for both true and recovered targets. Here, a successful detection (green circle) occurs when the estimated target is within one range cell, one azimuth bin and one Doppler bin of the ground truth (blue circle). More experiments in~\cite{mishra2016cognitive, cohen2016radarconmimo} demonstrate that the prototype performance is robust with SNRs dropping to as low as $-10$ dB and the time and spatial resolution are preserved by simulating couples of close targets in range, Doppler and azimuth.

\section{Conclusion  and Future Challenges}

In this paper, we reviewed several compressed radar systems that aim at reducing complexity while preserving parameter resolution. Throughout the review, we considered different popular radar systems, including pulse-Doppler and step frequency radars, as well as MIMO configurations. In particular, we showed that temporal, spectral and spatial compression can be implemented without decreasing Doppler, range and azimuth resolution. It has been shown that to recover these parameters for $L$ targets, the minimal number of required samples per pulse, minimal number of pulses and minimal number of channels are each equal to $2L$. These are determined by the actual number of degrees of freedom of the parameter estimation problem, governed by $L$, rather than a function of design parameters such as signal bandwidth, CPI or aperture. This is essential since the latter determine range, Doppler and azimuth resolution and are increased for higher performance. By breaking the traditional links between sampling rate, number of pulses and antennas on the one hand and parameter estimation on the other, increased performance may be achieved without increasing sampling and processing rates. 

An advantage of the Xampling system is that traditional radar processing algorithms can be easily adapted and applied directly to the sub-Nyquist samples. For example, clutter cancellation techniques have been implemented on the Xampling radar prototypes. These significantly enhance the performance of compressed radars without requiring reconstruction of Nyquist rate samples. In addition, while CS based methods traditionally do not perform well in the presence of large noise, since they inherently reduce SNR, Doppler focusing, applied to samples obtained using Xampling, enjoys an SNR improvement which scales linearly with the number of pulses, obtaining good detection at low SNRs.

An essential part of the approach adopted in this survey is the relation between the theoretical algorithms and practical hardware implementation, demonstrating real-time target detection from compressed samples, in the fast and slow time domains as well as in space. The prototypes presented here were built from off-the-shelf components paving the way to enabling commercial compressed radar systems. To this end, such hardware prototypes should be further extended to implement both radar transmitter and receiver systems, and deployed to be tested on real data. This would permit assessing their performance in real world conditions, including different types of noise, clutter and interference.

\bibliographystyle{IEEEtran}
\bibliography{IEEEabrv,CR_ref}

\end{document}